\documentclass{emulateapj}
 
\newcommand\lya{Ly$\alpha$}
\newcommand\ha{H$\alpha$}
\newcommand\gp{$g_{475}$}

\newcommand\ip{$i_{775}$}

\newcommand\zp{$z_{850}$}
\newcommand\bp{$B_{435}$}

\newcommand\vp{$V_{606}$}

\newcommand\idrops{$i$-dropouts }
\newcommand\idrop{$i$-dropout }
\newcommand\gdrops{$g$-dropouts }
\newcommand\gdrop{$g$-dropout }
\newcommand\vdrops{$V$-dropouts }
\newcommand\vdrop{$V$-dropout }
\newcommand\bdrop{$B$-dropout }
\def\Msun{M$_{\odot}$}

\slugcomment{Version 1.2, \today}
 
\shorttitle{Stellar masses at $z=4-6$}
\shortauthors{Overzier, R.A. et al.} 
 
\begin{document}
 
\title{Stellar Masses\altaffilmark{1} of Lyman Break Galaxies, Ly$\alpha$ Emitters and 
Radio Galaxies in Overdense Regions at $z=4-6$} 

 
\author{
Roderik A. Overzier\altaffilmark{2}, 
Xinwen Shu\altaffilmark{3,4}, 
Wei Zheng\altaffilmark{3},
Alessandro Rettura\altaffilmark{3},
Andrew Zirm\altaffilmark{3},
Rychard J. Bouwens\altaffilmark{5},
Holland Ford\altaffilmark{3},
Garth D. Illingworth\altaffilmark{5},
George K. Miley\altaffilmark{6},
Bram Venemans\altaffilmark{7},
Richard L. White\altaffilmark{8}
} 
\altaffiltext{1}{Based on observations with the NASA/ESA Hubble Space 
Telescope, obtained at the
Space Telescope Science Institute, which is operated by the Association of 
Universities of Research in Astronomy, Inc., under NASA contract NAS5-26555, 
and with the Spitzer Space Telescope, which is operated by the Jet Propulsion 
Laboratory, California Institute of Technology under a contract with NASA. 
}
\altaffiltext{2}{Max-Planck-Instit\"ut f\"ur Astrophysik, Karl-Schwarzschild-Strasse 1, D-85748 Garching, Germany.}
\altaffiltext{3}{Department of Physics and Astronomy, The Johns Hopkins University, Baltimore, MD 21218.}
\altaffiltext{4}{Center for Astrophysics, University of Science and Technology of China, Hefei, Anhui 230026, China.}
\altaffiltext{5}{Astronomy Department, University of California, Santa Cruz, CA 95064.}
\altaffiltext{6}{Leiden Observatory, University of Leiden, Postbus 9513, 2300 RA Leiden, The Netherlands.}
\altaffiltext{7}{European Southern Observatory, Karl-Schwarzschild-Strasse 2, D-85748 Garching, Germany.)}
\altaffiltext{8}{Space Telescope Science Institute, 3700 San Martin Drive, Baltimore, MD 21218.}

\email{overzier@mpa-garching.mpg.de}
 
\begin{abstract}
  We present new information on galaxies in the vicinity of luminous radio galaxies and quasars at $z\simeq$4, 5, and 6.  These fields were previously found to contain overdensities of Lyman Break Galaxies (LBGs) or spectroscopic Ly$\alpha$ emitters, which were interpreted as evidence for clusters-in-formation (``protoclusters''). We use HST and Spitzer data to infer stellar masses from stellar synthesis models calibrated against the Millennium Run simulations, and contrast our results with large samples of LBGs in more average environments as probed by the Great Observatories Origins Deep Survey (GOODS).  The following results were obtained. First, LBGs in both overdense regions and in the field at $z=4-5$ lie on a very similar sequence in a $z^\prime$--[3.6] versus 3.6$\mu$m color-magnitude diagram. This is interpreted as a sequence in
  stellar mass ($M_*\sim10^{9}$--$10^{11}$ $M_\odot$) in which
  galaxies become increasingly red due to dust and age as their star formation rate (SFR)
  increases, while their specific SFR stays constant. Second, the two radio galaxies are among the most massive objects ($M_*\sim10^{11}$ $M_\odot$) known to exist at $z\simeq4-5$, and are extremely rare based on the low number density of such objects as estimated from the $\sim$25$\times$ larger area GOODS survey.  We suggest that the presence of the massive (radio) galaxies and associated supermassive black holes has been boosted through rapid accretion of gas or merging inside overdense regions. Third, the total stellar mass found in the $z=4$ protocluster TN1338 accounts for $<$30\% of the stellar mass on the cluster red sequence expected to have formed at $z\gtrsim4$, based on a comparison with the massive X-ray cluster Cl1252 at $z=1.2$.  Although future near-infrared observations should determine
  whether any massive galaxies are currently being missed by our
  UV/\lya\ selections, one possible explanation for this mass
  difference is that TN1338 evolves into a smaller cluster than
  Cl1252.  This raises the interesting question of whether the most
  massive protocluster regions at $z>4$ remain yet to be discovered.
\end{abstract}
 
 
\section{INTRODUCTION} 

Galaxy clusters formed in high-density regions of dark matter that
collapsed earlier than surrounding low-density regions
\citep{kaiser84,springel05}, and models predict that the galaxies that
formed inside these high-density regions also may have formed earlier
or evolved at an increased rate compared to their surroundings
\citep{kauffmann95,benson01,thomas05,delucia06}. The mass
functions of clusters at low and high redshift can significantly improve
current estimates of the cosmological parameters
\citep{vikhlinin09}.  Clusters of galaxies are thus important 
not only for the formation of dark matter
structure, but also for studying processes related to galaxy
formation, such as the history of gas accretion and depletion, the
feedback of supernovae and active galactic nuclei (AGN), the 
enrichment of the intracluster medium, and the spectra-morphological transitions of galaxies.

In the local Universe, the galaxies in the centers of clusters inhibit
a distinct location in the color-magnitude diagram (CMD). This cluster
``red sequence'' consists of predominantly spheroidal and lenticular
galaxies having old stellar populations and high stellar masses, and
usually includes the brightest cluster galaxy. In order to understand
how and when this relation formed, it is important to study clusters
at high redshifts.  Traditional surveys have uncovered
clusters at redshifts as high as $z\sim1.5$
either by searching for galaxies on
the characteristic red sequence or by searching for hot cluster gas in
the X-rays. A red sequence of morphological early-type galaxies is
prominently present in even the most distant clusters, and the small
scatter in galaxy colors around the sequence has been used to infer
formation redshifts of $\sim2-5$
\citep[e.g.][]{ellis97,vandokkum00,stanford02,stanford05,blakeslee03a,blakeslee06,holden05,postman05,mullis05,mei06,rettura08}.

The epoch corresponding to the assembly of massive cluster galaxies  
is presumed to be marked by a violent stellar mass
build-up. In order to target this epoch, various techniques have been
used. One technique is based on the hypothesis that at very early
times the most massive galaxies trace the highest density regions and can
thus be used as ``lighthouses'' to pinpoint possible
``protoclusters''.  Radio galaxies (RGs) are among the largest and
most massive galaxies at $z=2-5$ \citep[$M_*\sim10^{10-12}$ M$_\odot$,
  e.g.][this paper]{pentericci97,rocca04,seymour07,hatch09}, while the most
luminous high redshift quasars (QSOs) have some of the largest black
hole masses inferred \citep[$M_{BH}\sim10^{8-10}$ M$_\odot$,
  e.g.][]{jiang07,kurk07}.  
Overdensities (of the order of a few) of star-forming galaxies
suggest that at least some radio
galaxies are associated with rare peaks in the large-scale structure
\citep[e.g.][]{pentericci00,miley04,venemans02,venemans07}.
Similarly, the number counts of faint $i$-dropouts towards two
$z\sim6$ QSOs \citep{stiavelli05,zheng06,kim08} also indicate a
possible enhancement \citep[see][for a detailed
  discussion]{overzier09}. Similar overdensities have also been found in
``random'' fields at $z\simeq2-6$
\citep[e.g.][]{steidel98,steidel05,shimasaku03,ouchi05,ota08}, and are consistent with being progenitors of
present-day clusters.

\citet{steidel05} found an increase in the stellar masses and ages for
star-forming galaxies in a protocluster at $z=2.3$ compared to the surrounding
field. In \citet{zirm08} we presented evidence for a 
population of relatively evolved galaxies near a radio galaxy at
$z=2.2$ \citep[see also][]{kodama07}, and
\citet{kurk04} found evidence for different spatial
distributions of blue and red galaxies. Although \citet{venemans07}
found no differences in the physical properties of LAEs in
protoclusters compared to the field, protoclusters tend to have an
elevated AGN fraction
\citep[e.g.][]{pentericci02,venemans07,lehmer08}. \citet{lehmer08}
found an enhancement in the stellar masses of LBGs in
the protocluster SSA22 at $z=3.1$ \citep{steidel98}, while
\citet{peter07} and \citet{overzier08} did not find any morphological
differences between UV-selected galaxies in protoclusters compared to
the field using data from the Advanced Camera for Surveys
(ACS) on-board the {\it Hubble Space Telescope} (HST).

The ACS, in particular, has enabled the selection of large samples of
faint UV ``dropout'' galaxies at $z>4$ \citep[][]{giavalisco04,bouwens04,bouwens07}. Combined with
the great sensitivity in the mid-infrared provided by the
Infrared Array Camera (IRAC) on the {\it Spitzer Space Telescope}
(SST), we can study the stellar populations of those galaxies
\citep[][]{yan05,labbe06,eyles07,stark07,stark09,verma07,yabe08}. While
observations in the UV probe the (dust attenuated) light from young,
massive stars, the rest-frame optical light
offers a more direct probe of the total stellar mass.  Results at
$z\sim2-3$ imply that some LBGs and near-infrared selected galaxies
were already quite massive at $z\gtrsim5$ implying rapid evolution
\citep[e.g.][]{shapley01,papovich01,yan05,papovich06,wuyts07}.
Recently, \citet{stark09} found that the $z>4$ UV luminosity function (LF)
is dominated by episodic star
formation ($\le$500 Myr on average) from systems that reached their
high UV luminosities in short times ($<$300 Myr), consistent with the
decline in the UV LF \citep{bouwens07}. Galaxies having relatively
high equivalent widths of \lya\ emission (the so-called
``\lya-emitters'') have received attention (partly due to the relative
ease of their detections out to the highest redshifts) concerning the
question whether they mark a particular phase in the evolution of
typical starbursts
\citep[e.g.][]{shapley03,gawiser06,pentericci07,pentericci09,ouchi08}.
  
In this paper we will derive stellar masses, as inferred from the
rest-frame optical fluxes, for a sample of LBGs, LAEs and
RGs\footnote{Stellar mass estimates for the radio galaxies
  TNJ1338--1942 and TNJ0924--2201 are also given in
  \citet{seymour07}. Our analysis differs in that it is based on our
  deeper data and uses more sophisticated photometry. However, our
  results are consistent.}  in two protoclusters: ``TN1338'' at $z\simeq4.1$ \citep{venemans02,miley04,zirm05,overzier08}, and ``TN0924'' at $z\simeq5.2$ \citep{venemans04,overzier06}. For completeness, we also present a small sample of \idrops near two quasars at $z\sim6$ studied by \citet{stiavelli05} and \citet{zheng06}. The target fields are listed in Table \ref{tab:samples}.  Comparison with large samples from the GOODS survey \citep{giavalisco04} will enable us to compare the overdense regions with more average environments. While the data presented here cannot compete in terms of either depth or area with other studies in the literature \citep[e.g.][]{stark09}, our data is unique in the sense that we cover a few sightlines towards rare, overdense regions and luminous AGN. Specifically, we will address the following questions:\\ 
(1) How do the stellar masses of radio galaxies, LBGs and LAEs compare?\\
(2) Do galaxies in overdense regions at high redshift have distinct physical properties?\\ 
(3) Is the total amount of stellar mass detected in protoclusters consistent with the stellar mass in galaxies on the low redshift cluster red sequence?\\
(4) Have we missed a population of obscured or quiescent galaxies not selected by our UV/\lya\ selections?

The structure of this paper is as follows: The samples
and our measurement techniques are described in \S2. In \S3 we present
the rest-frame UV-optical color-magnitude diagrams at $z\simeq4,5,6$
and estimate stellar mass distributions based on simple model
comparisons. We discuss our results in \S4. All magnitudes are in the
AB system. For comparison with other studies, it is convenient to
specify that the characteristic luminosity $L_{z=3}^*$ at $z\sim3$
\citep[][i.e., $M_{1700,AB}=-21.07$]{steidel99} corresponds to
$z$-band magnitudes of $\sim$24.9, 25.3 and 25.6 at $z$$\sim$4, 5 and 6,
respectively.  We adopt the following cosmological parameters:
$\Omega_M=0.27$, $\Omega_\Lambda=0.73$, and $H_0=73$ km s$^{-1}$
Mpc$^{-1}$.
 
\section{SAMPLES, DATA AND MEASUREMENTS}

\subsection{Primary Samples}
 
\subsubsection{Radio galaxy fields}

\noindent
This paper is based on various samples of galaxies in overdense
regions selected as part of previous works (see \S1).  For the two
radio galaxy fields we make use of large samples of LBGs and LAEs
selected from our previous HST/ACS imaging and \lya\ surveys.

In the field TN1338 we will focus on a sample of LAEs and LBGs
selected near the radio galaxy TN J1338--1942 at $z=4.11$.  The
  field covered with both HST and Spitzer contains 12 \lya\ emitters
  from \citet{venemans07}: these LAEs are part of a larger structure
  that is $\approx5\pm1$ times more dense compared to average
  expectations, and has a velocity dispersion of $\approx300$ km
  s$^{-1}$ centered on the radio galaxy. The LBG sample consists of 66
  \gp-band ($g$) dropouts brighter than \zp$=$27.0 mag. As detailed in
  \citet{overzier08}, the photometric redshift distribution peaks at
  $z=4.1$, and has a width of $\Delta z\sim0.5$ (FWHM). About half of
  the LAEs ($|z-\Delta z|<0.03$) are also selected as part of the
  dropout sample, indicating that the true width of the redshift
  distribution of the dropouts could be narrower than suggested by the
  photo-$z$ analysis given its fairly limited accuracy. Unfortunately,
  the redshifts of most of the dropouts are difficult to determine
  spectroscopically due to their faintness and because all high
  equivalent width \lya\ sources have already been
  identified. Statistical studies indicate that the ratio of
  \lya-faint to \lya-bright LBGs in a UV flux-limited sample is about
  3:1 \citep[e.g.][]{shapley03}. The population of
  $g$-dropouts represents an overdensity at the $>5\sigma$
  significance with respect to the average field as estimated from the
  GOODS survey, consistent with a substantial fraction lying near the
  redshift of the radio galaxy and its associated \lya\ emitters
  \citep{overzier08}. The LBG/LAE samples are listed in Tables
\ref{tab:lbgz4} and \ref{tab:laesz4}.

In the field TN0924 we make use of our samples of LAEs and LBGs
selected near the radio galaxy TN J0924--2201 at $z=5.20$. This galaxy
is the highest redshift radio galaxy known \citep{vanbreugel99}. 
  The radio galaxy has six spectroscopically confirmed LAE companions
  \citep[$|z-\Delta z|<0.015$;][]{venemans04}, four of which are in
  the field covered by HST and Spitzer. The LBG sample consists of 23
  \vp-band ($V$) dropouts brighter than \zp$=$26.5 mag. Analogous to
  the TN1338 sample, the $V$-dropout selection was tuned to the
  selection of candidate LBGs near the redshift of the radio galaxy
  \citep{overzier06}.  The TN0924 field contains about twice as many
  $V$-dropouts compared to the average number found in GOODS, with a
  $\sim$1\% chance of finding a similarly high number also based on a
  comparison with GOODS. Their photometric redshift distribution has a
  width of $\Delta z\sim0.5$ (FWHM), but its true width could be
  narrower based on the same arguments as given above. Two of the LAEs
  are also selected as dropouts, implying that, out of all candidates
  selected, the actual number of LBGs physically associated with the
  radio galaxy and other LAEs may not be higher than $\sim8$ for a
  \lya\ faint-to-bright ratio of 3:1 \citep[see][for
    discussion]{overzier06}. The samples are listed in Table
\ref{tab:lbgz5}.

\subsubsection{Quasar Fields}

For the {\it QSO field} SDSS0836 we make use of a sample of
$i$-dropouts that was selected in the
3.4\arcmin$\times$3.4\arcmin\ HST/ACS pointing towards the radio-loud
QSO SDSS J0836+0054 at $z=5.82$. This QSO is the highest redshift
radio-loud quasar known. \citet[][]{zheng06} found 7 dropouts within a
5 arcmin$^2$ projected region near the QSO, having colors consistent
with them being at $z\approx5.8$. In addition, we obtained archival
ACS observations (GO9777, PI: Stiavelli) of the field towards QSO SDSS
J1030+0524 at $z=6.28$ ({\it QSO field} SDSS1030). The data was
processed with the pipeline APSIS \citep{blakeslee03b}, and we
performed our own selection of \idrops using SExtractor
\citep{bertin96} and using the same criteria as in \citet{zheng06}: $i
- z > 1.3$ (mag\_iso), $z\le26.5$ mag, and a SExtractor star/galaxy
index $< 0.8$.  We visually inspected the ACS images and excluded
candidates near the detector edges. Because of our own processing of
the data as well as more stringent selection criteria, the number of
our candidates (5) in the SDSS1030 field is somewhat smaller than that
given previously in the literature \citep{stiavelli05,kim08}.  The two
samples of $i$-dropouts in QSO fields are listed in Table
\ref{tab:qsoz6}.

\subsection{GOODS Comparison Data}

Our comparison data is based on the large dropout samples from
  \citet{stark07} and \citet{stark09}, who performed a detailed study
  of the high redshift LBG populations in the $\sim320$ arcmin$^2$
  GOODS survey \citep{giavalisco04}. The parent samples from Stark et
  al. consist of 2443, 506 and 137 $B$-, $V$-, and $i$-dropouts to a
  limiting optical magnitude of 28, of which $\sim$35\% was found to
  be sufficiently isolated in the IRAC images for accurate
  photometry. The selection criteria for these samples are equivalent
  to those used for our protocluster fields. Details on the selection
  and the HST and Spitzer photometry of the GOODS samples can be found
  elsewhere \citep{stark07,stark09}.

For comparison with our TN1338 field, we will make use of their large
sample of $z\approx4$ $B$-dropouts, while for TN0924 we will use the
sample of $z\approx5$ $V$-dropouts. For the two QSO fields at
$z\approx6$ we use a small sample of $i$-dropouts also from
GOODS. After applying appropriate magnitude cuts, we are left with
420, 122 and 42 comparison objects, respectively. Because the
overdense fields listed above have all been previously demonstrated to
be overdense relative to the much larger GOODS survey, the latter is
expected to provide a good handle on the LBG population in more
average environments.

\subsection{Millennium Run Simulations Comparison Samples}
\label{sec:mrsample}

In order to test our strategy for determining stellar masses based on
a comparison between simple SED models and the observed magnitudes, we
have selected a large sample of mock galaxies simulated as part of the
Millennium Run (MR) Simulations project \citep{springel05}. In brief,
galaxies were modeled by applying semi-analytic model (SAM)
prescriptions of galaxy formation to the dark matter halo merger trees
from the MR. The SAM includes processes such as gas cooling, star
formation, reionization heating, and feedback from supernova and AGN
that are coupled to the stellar synthesis model library from BC03 in
order to calculate the resulting spectrum of each galaxy in the
simulations as its evolution is tracked across discrete ``snapshots''
in time between $z=127$ and $z=0$ \citep[see][for
  details]{croton06,delucia07,guo08}. We use the mock galaxy
catalogues `delucia2006a' from \citet{delucia07} and randomly select 1
million galaxies each at $z\approx4$ and $z\approx5$. Previous studies
have shown that the simulations are able to reproduce the observed
properties of galaxies at these redshifts relatively well
\citep[e.g.][]{guo08,overzier09}. The large size of the mock galaxy
catalogues selected ensures that we sample a wide range in SFHs, ages, masses, dust
attenuations, metallicities and large-scale environments.

\subsection{IRAC Imaging}

\begin{figure}[t]
\includegraphics[width=\columnwidth]{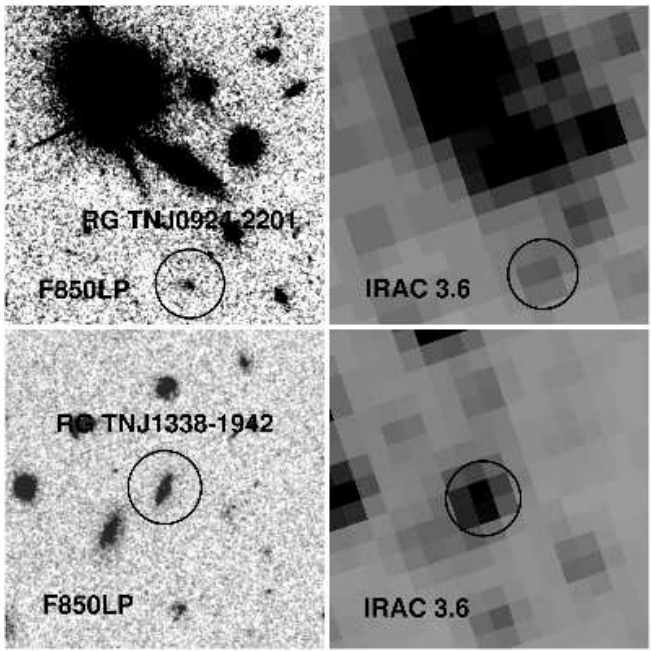} 
\caption{Panels show the ACS \zp-band and IRAC 3.6$\mu$m images of the
  radio galaxies TN1338 at $z=4.1$ (bottom panels) and TN0924 at
  $z=5.2$ (top panels). The images measure
  20\arcsec$\times$20\arcsec. The ACS images were convolved with a
  Gaussian kernel of $0\farcs1$ (FWHM). The IRAC photometry was
  obtained by deblending the radio galaxies and their neighbouring
  sources using GALFIT (see \S2 for details).}
\label{fig:stamps}
\end{figure}

Follow-up infrared images for the fields TN0924 and SDSS0836 were
obtained with the IRAC on the SST (PID 20749, Zheng 2006). We
retrieved archival IRAC data of TN1338 (PID 17, Fazio 2004) and
SDSS1030 (PID 30873, Labb\'e 2007).  The IRAC camera includes imaging
data at 3.6, 4.5, 5.8, and 8.0 $\mu$m (channels 1-4), each with a
$5\farcm2\times5\farcm2$ field and a pixel size of $\sim1\farcs22$. We
concentrate here on the shortest-wavelength {\it Spitzer} image
(channels 1 and 2) for the field at $z\sim4$ (TN1338), and on the
3.6$\mu$m images for the higher redshift fields. The exposure times at
3.6\micron\ are 17 ksec (TN0924 \& SDSS0836) and 10 ksec (TN1338 \&
SDSS1030).  For our photometry, we used the pipeline-processed IRAC
images at the `post-basic calibrated data' (PBCD) stage.  The main
steps in the pipeline include dark current subtraction, flat fielding,
flux calibration (in units of MJy sr$^{-1}$), geometric distortion
correction and mosaicking of the individual frames.  Sensitivities of
each image were estimated from the 2 $\sigma$ standard deviation in
the flux distribution measured in randomly placed 3\arcsec\ diameter
apertures.  The 2$\sigma$ limiting AB magnitudes are 25.5 (TN0924),
25.2 (J0836), 25.2 (TN1338), and 25.2 (J1030) mag in the 3.6 $\mu$m
channel, and 25.0 (TN1338) and 24.6 (J1030) mag in the 4.5 $\mu$m
channel.

\subsection{IRAC Source Photometry using GALFIT}

The IRAC images suffer from overcrowding due to its relatively large
point spread function (PSF) of $\sim1\farcs5$ (FWHM) and great
sensitivity.  To address this issue, we use a deblending technique
whereby contaminating neighbors are subtracted using GALFIT
\citep{peng02} by performing a fit to the objects of our interest and
all their close neighbours simultaneously.  GALFIT constructs a
two-dimensional model of the data according to specified input
parameters (e.g., positions, magnitude, effective radius, axis ratio,
and position angle), performs a convolution with the instrument PSF,
and fits the result to the data through an iterative
$\chi^2$-minimization process.  The IRAC PSF is obtained directly from
the images by stacking 10 bright, isolated point sources.  We use the
higher resolution ACS $z$-band image as the reference for the initial
GALFIT input parameters.  GALFIT was then applied to the IRAC
images. Two versions of the fitting process were carried out. First,
all input parameters as determined from the $z$-band image were held
fixed, except the magnitudes.  The resulting $\chi^2$ value and
residual image was then visually inspected to determine the goodness
of fit. When the best possible fit was not achieved, we repeated the
GALFIT procedure, but this time allowing all input parameters to
vary. Each object was individually and interactively processed until
the most successful fit was achieved, while avoiding the
oversubtraction of the flux contribution from the neighboring sources.
Following this, the two-dimensional models of the light profiles of
the LBGs/LAEs and nearby contaminating sources were deblended. This
deblending process works quite well for determining the intrinsic
fluxes of most of LBGs/LAEs.  Those sources for which GALFIT failed to
satisfactorily deblend the emission of LBGs/LAEs from the neighbour
sources were removed from the sample.  This leaves a total of 60
sources (38 LBGs and three LAEs at $z\simeq4.1$, 15 LBGs at $z\simeq
5.2$ and 4 LBGs at $z\sim 6$), for which reliable photometry from
GALFIT was obtainable.

The measured 3.6 (4.5) $\mu$m magnitudes for the four fields are
listed in Tables 1--4, along with the classifications of galaxies for
their detections and degree of confusion, as (1) isolated and
detected; (2) isolated but undetected; (3) confused (4) heavily
blended.  In Fig. \ref{fig:stamps} we show the ACS \zp\ and IRAC
3.6$\mu$m images of the two bright radio galaxies for an illustration
of the data quality and blending issues.

\subsection{Emission line contamination corrections}
\label{sec:ha}

Because the IRAC 3.6$\mu$m flux of the radio galaxy TN J1338--1942 is
significantly contaminated by \ha\ flux from its large emission line
halo that is typical of RGs, we have used the emission-line free flux
at 4.5$\mu$m and a small model extrapolation based on the typical
[3.6]--[4.5] colors of star-forming galaxies at $z\approx4$ to
estimate the true continuum magnitude at 3.6$\mu$m as follows. The
measured [3.6]--[4.5] color was --0.65 mag (see Table 1),
while a comparison with stellar population models (see below) in
4.5$\mu$m vs. $z$--[4.5] space would predict a color of
[3.6]--[4.5]$\sim$0--0.2 mag. We thus assume that
\ha\ contributes about $0.75\pm0.10$ magnitude to the 3.6$\mu$m
measurement of the radio galaxy, and we will apply the correction to
all subsequent figures and tables. The 3.6$\mu$m flux of the radio
galaxy TN J0924--2201 is not affected by \ha\ given its higher
redshift of $z=5.2$.

While the 3.6$\mu$m flux of $z\sim4$ star-forming galaxies such as
LBGs and LAEs is expected to have an enhancement due to \ha\ as well,
we do not correct for this effect as we can assume that galaxies in
both GOODS and in the protocluster fields will be affected in a
similar way. \citet{stark07} estimate that \ha\ could in principle
increase the broad band flux by $\sim10-20$\%.

\begin{figure}[t]
\begin{center}
\includegraphics[width=\columnwidth]{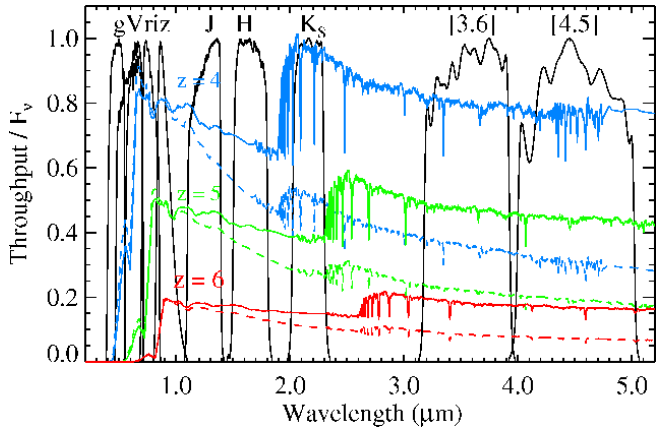} 
\end{center}
\caption{Transmission curves of the relevant
    optical (HST/ACS), near-infrared, and infrared (IRAC)
    filters. We overplot several example spectral energy distributions modeled
    using BC03. Lines show the expectation for an exponentially
    declining star formation history with $\tau=100$ Myr at ages of 10
    (dashed lines) and 100 Myr (solid lines), and redshifted to
    $z=4.1$ (blue lines), $z=5.2$ (green lines), and $z=6.0$ (red
    lines). A small attenuation of $E(B-V)=0.1$ magnitude was modeled using
    the prescription of \citet{calzetti01} and applied to each of the tracks. 
 \label{fig:filters}}
\end{figure}

\subsection{Estimating stellar masses}
\label{sec:massmethod}

\subsubsection{Rationale}

In order to answer the question of whether there are any differences
between the different types of galaxies across the CMD and across
environment, we will seek to determine estimates of (primarily)
stellar masses based on a set of $z$--[3.6] versus 3.6$\mu$m
CMDs. Only recently has it become feasible to attempt an
interpretation of the observed magnitudes of the relatively faint
galaxy populations at $z\simeq3-6$ in terms of stellar masses, ages,
dust and star formation histories based on detailed fitting of SEDs
using model libraries (see references given in \S1 and therein).
While there is ongoing debate about the correctness of the treatment
of different stellar evolutionary phases in the model libraries
\citep[e.g.][]{fioc97,bruzual03,maraston06,rettura06,kannappan07,eminian08}, they at least
allow for an assessment of the {\it comparative} properties among
samples across redshift or across the CMD, provided that these samples
are sufficiently `simple' to model and that the adopting of a faulty
model does not lead to catastrophic systematic effects across the
various samples. At the least, this requires highly accurate
photometry in all of the rest-UV, optical and infrared, and such
studies have therefore been limited to a small set of well-studied
fields.

While our ACS imaging is of comparable depth to, e.g., GOODS, we lack
sufficiently deep data in the observed near-infrared. Our IRAC imaging
provides a relatively good match to the ACS imaging (most UV-selected
sources detected), but it is not as deep as some of the large field
surveys. Therefore, we do not believe that a detailed SED fitting is
warranted by our data, nor is it necessary for the limited questions
we wish to ask.  Instead, as we will show below, we can obtain
sufficiently accurate information on stellar masses based on the
comparison with simple baseline models rather than SED fitting
(\S\ref{sec:baseline}). We will test our method by making a comparison
with the properties of mock galaxies selected in a large cosmological
simulation (\S\ref{sec:mrtest}).

\subsubsection{Comparison with Stellar Synthesis Models}
\label{sec:baseline}

\begin{figure}[t]
\includegraphics[width=\columnwidth]{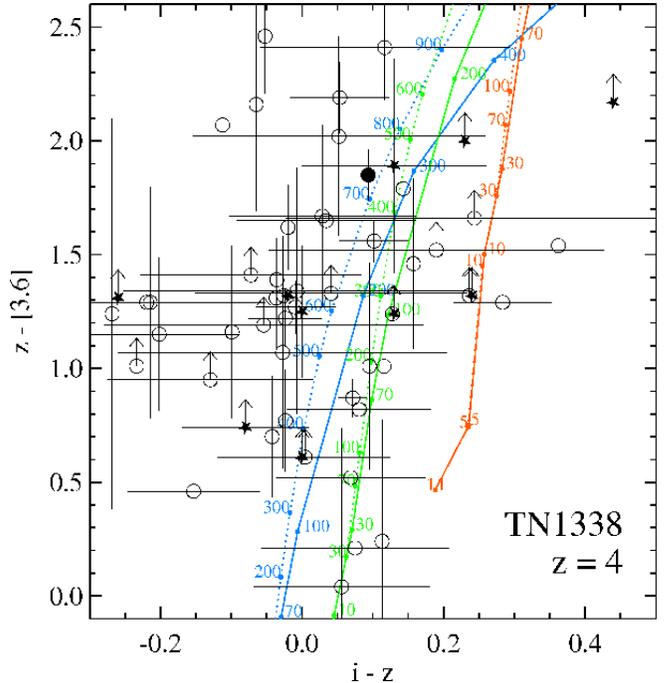}
\caption{The $i$--$z$ versus $z$--[3.6] color-color diagram for
  the TN1338 field. Symbols indicate the radio galaxy (large filled
  circle), LBGs (open circles) and LAEs (filled stars). Tracks
  indicate the colors of models having $\tau=30$ Myr (dotted lines)
  and $\tau=200$ Myr (solid lines) with ages in Myr indicated along
  the tracks. Attenuations of $E(B-V)=0.0,0.15,0.4$ mag are shown in
  blue, green and red, respectively. Most
  objects do not require a large amount of reddening
  due to dust.}
  \label{fig:iz}
\end{figure}

As detailed in \citet{stark09}, at $z\gtrsim 4$ the 3.6$\mu$m channel
still probes only as red as the rest-frame optical, where the
contribution from TP-AGB stars is much less severe than in the
near-infrared. As a result, the stellar mass estimates based on
``BC03'' \citep{bruzual03} models are expected to be higher by at most
$\simeq$30\% when compared to the newer generation of models that
include this phase \citep{bruzual07}, and the effect of AGB stars
becomes less severe with increasing redshift and metallicity with
respect to subsolar metallicity models at $z\sim4$. Furthermore, the
SED fitting appears to be relatively insensitive to whether redshifts
are being fixed or allowed to vary within their $\Delta z=\pm0.25$
uncertainty distributions typical for dropout samples.
 
In order to interpret the data in the CMDs that will be presented in
\S\ref{sec:results}, we have selected a set of BC03 models that we
will use for a simple baseline comparison with our data. The models
are based on the best-fit ages as obtained from pan-chromatic SED fits
performed by \citet{stark09}, and the typical attenuations due to dust
as found by \citet{bouwens09}. Both studies are based on large samples
of dropout galaxies at $4\lesssim z\lesssim6$ for which deep
photometry is available. We choose an exponentially declining star
formation history with an $e$-folding timescale $\tau$ of 100 Myr. We
consider a range of ages from 100 to 300 Myr, straddling the best-fit
age of $\sim$200 Myr as found by \citet{stark09}. We will also
consider three levels of attenuation by dust in terms of $E(B-V)=$0.0
mag (no dust), 0.15 mag (moderate extinction) and 0.3 magnitude
(dusty), modeled using the recipes for starburst galaxies given in
\citet{calzetti01}. All models have a metallicity similar to that of
the Large Magellanic Cloud. We convert the models to the appropriate
redshift, and calculate colors and magnitudes by convolving the
spectral energy distributions with the HST and Spitzer filter
transmission curves (Fig. \ref{fig:filters}). Given an age and an
attenuation, an estimate for the stellar mass of each object is
obtained by determining the absolute mass scale of the model required
in order to reproduce the 3.6$\mu$m magnitude observed. Motivated by a
recent study by \citet{bouwens09} that finds clear evidence of
correlations between the UV color and absolute UV luminosity of $B$-
and $V-$dropout galaxies, we will also investigate a model in which
the dust and age are allowed to vary as a function of absolute
magnitude.

These simple model comparisons show that while the 3.6$\mu$m magnitude
is an approximate gauge of stellar mass at each of the redshifts, the
$z$--[3.6] color is affected by significant degeneracies between
age, dust and SFH that can only be satisfactorily addressed using
deeper data and a wider set of bands than currently available for our
fields.  However, since this paper mainly focuses on general trends in
stellar mass for different populations, our main results will not be
significantly affected by this issue. Nonetheless, for the TN1338
field we can be a little more specific and try to get a better handle
on the presence of dust. The $i$--$z$ UV continuum color reacts more
strongly to reddening due to dust than the $z$--[3.6] color. This
is illustrated in Fig. \ref{fig:iz} where we show the TN1338 $i$--$z$
vs. $z$--[3.6] color-color diagram compared to models with three
different attenuations of 0.0 (blue lines), 0.15 (green lines) and 0.4
mag (red lines). The symbols indicate LBGs (open circles), LAEs
(filled stars), and the RG (filled circle). Most objects are very blue
in the UV ($i$--$z\approx0.0$ mag) implying a small amount of
attenuation at most. In other words, strongly reddened galaxies do not
appear to make an important contribution. This is not surprising since
not only is our UV selection biased against very dusty objects, recent
evidence suggests that dusty objects become increasingly rare at
$z\gtrsim4$ \citep[e.g.][Sect. \ref{sec:raredust} of this
  paper]{mclure06,stark09,bouwens09}. 

\subsubsection{Comparison with Mock Galaxies in the Millennium Run Simulations}
\label{sec:mrtest}

\begin{figure}[t]
\begin{center}
\includegraphics[width=\columnwidth]{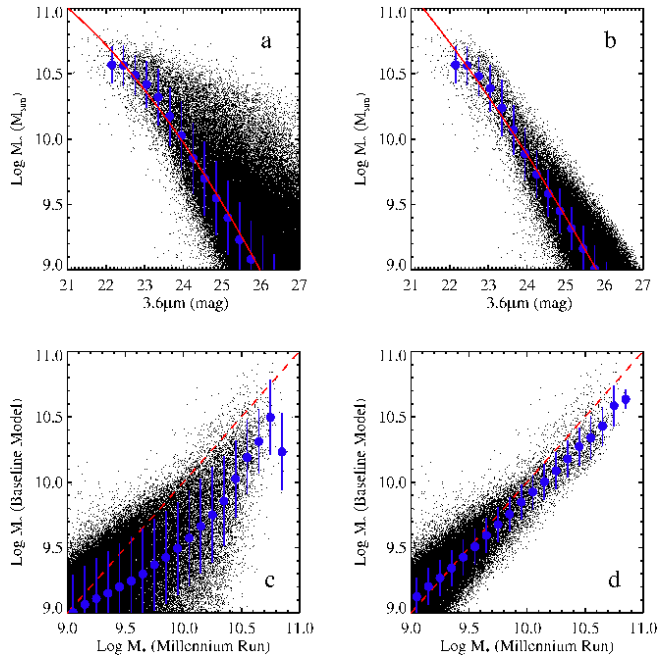}
\end{center}
\caption{The accuracy of our method of retrieving stellar masses based
  on simple SED assumptions, tested against one million mock galaxies
  selected from the Millennium Run (MR) cosmological simulations. Top
  panels show the relation between stellar masses and 3.6$\mu$m
  magnitudes for the entire MR sample (panel a, left) and for a
  cleaned subset of galaxies most similar to our UV-selected samples
  (panel b, right). Blue points indicate the median values and the
  1$\sigma$ standard deviation in bins of $\Delta m=0.3$ mag.  A
  second-order polynomial fit to the correlation shown in panel b is
  $\mathrm{log}M_* [M_\odot]=11.51+0.32m_{3.6}-0.016m_{3.6}^2$ (red
  line). In the bottom panels, we demonstrate how accurately we can
  derive the true stellar masses from the observed 3.6$\mu$m
  magnitudes under the assumption of a fixed baseline SED
  characterized by $\tau=100$ Myr, $t=200$ Myr, $E(B-V)=0.15$
  mag, and LMC metallicity. Despite the inevitable mismatches between
  the baseline model and the true properties of the simulated
  galaxies, there is still a very good correlation between the true
  mass and the mass inferred using the baseline model. The one-to-one
  correlation is indicated by the red dashed lines. There is a small
  systematic offset in the sense that for more massive galaxies we
  underestimate their true mass based on the baseline SED model (by
  about $\sim$0.3 dex at $M_*\sim10^{11}$ $M_\odot$, see panel d for
  the sample most similar to our UV selection).}
\label{fig:mrtest}
\end{figure}

Above we have outlined our basic strategy of using simple baseline
models to convert 3.6$\mu$m fluxes into stellar mass estimates. In
order to test this we use our large sample of mock galaxies from the
Millennium Run simulations (see Sect. \ref{sec:mrsample}). In the
top-left panel of Fig. \ref{fig:mrtest} ({\it panel a}) we plot the
stellar mass against the observed-frame 3.6$\mu$m magnitude for each
of the mock galaxies at $z=4$. The 3.6$\mu$m magnitude is clearly
correlated with the stellar mass, indicating that the flux at
rest-frame $\sim$7000\AA\ is a good tracer of the mass. Next, we limit
the mock sample to only those galaxies that are most similar to the
galaxies in our UV-selected LBG samples by requiring relatively little
dust ($E(B-V)<0.3$ mag, see Fig. \ref{fig:iz}) and current star
formation activity at a rate of $>$1 $M_\odot$ yr$^{-1}$. The result
is shown in the top-right panel ({\it b}). The scatter around the
correlation between $M_*$ and 3.6$\mu$m is significantly reduced. We
see that a range in the 3.6$\mu$m magnitude from 26 to 21 mag
corresponds to a range in stellar mass from $10^9$ to $10^{11}$
$M_\odot$. A second-order polynomial fit to the data yields
$\mathrm{log}M_* [M_\odot]=11.51+0.32m_{3.6}-0.016m_{3.6}^2$ (red line
in {\it panel b})

As stated earlier there are ambiguities between SFH, age and dust that
cannot be solved for based on our current data.  However, we can also
use the MR simulations to test how well we can retrieve stellar masses
when assuming that each galaxy can be approximated by a single one of
our simple baseline SEDs from Sect. \ref{sec:baseline} (e.g.,
$\tau=100$ Myr, $t=200$ Myr, $E(B-V)=0.15$ mag) for each
3.6$\mu$m magnitude. The result is shown in the bottom panels of
Fig. \ref{fig:mrtest}, again for all ({\it left panel, c}) and for the
``cleaned'' subsample ({\it right panel, d}). Despite the inevitable
mismatches between the baseline model and the true properties of the
simulated galaxies, there is a very good correlation between the
true mass and the mass inferred using the baseline model. We do
however see a systematic offset in the sense that for more massive
galaxies we underestimate their true mass based on the baseline SED
model (by about $\sim$0.3 dex at $M_*\sim10^{11}$ $M_\odot$). This can
be understood because more massive objects have higher SFRs and have
more dust in the simulations. Therefore, they also will have more dust
compared to our baseline model so their 3.6$\mu$m flux is lower and
hence we get a lower mass using the baseline model conversion.

The results obtained for our $z=5$ sample are very similar, but are
not shown here for brevity. In summary, these simulations demonstrate
that our approach of using simple baseline models to infer stellar
masses from the 3.6$\mu$m data is justified.
 
\begin{figure*}[t]
\begin{center} 
\includegraphics[width=1.5\columnwidth]{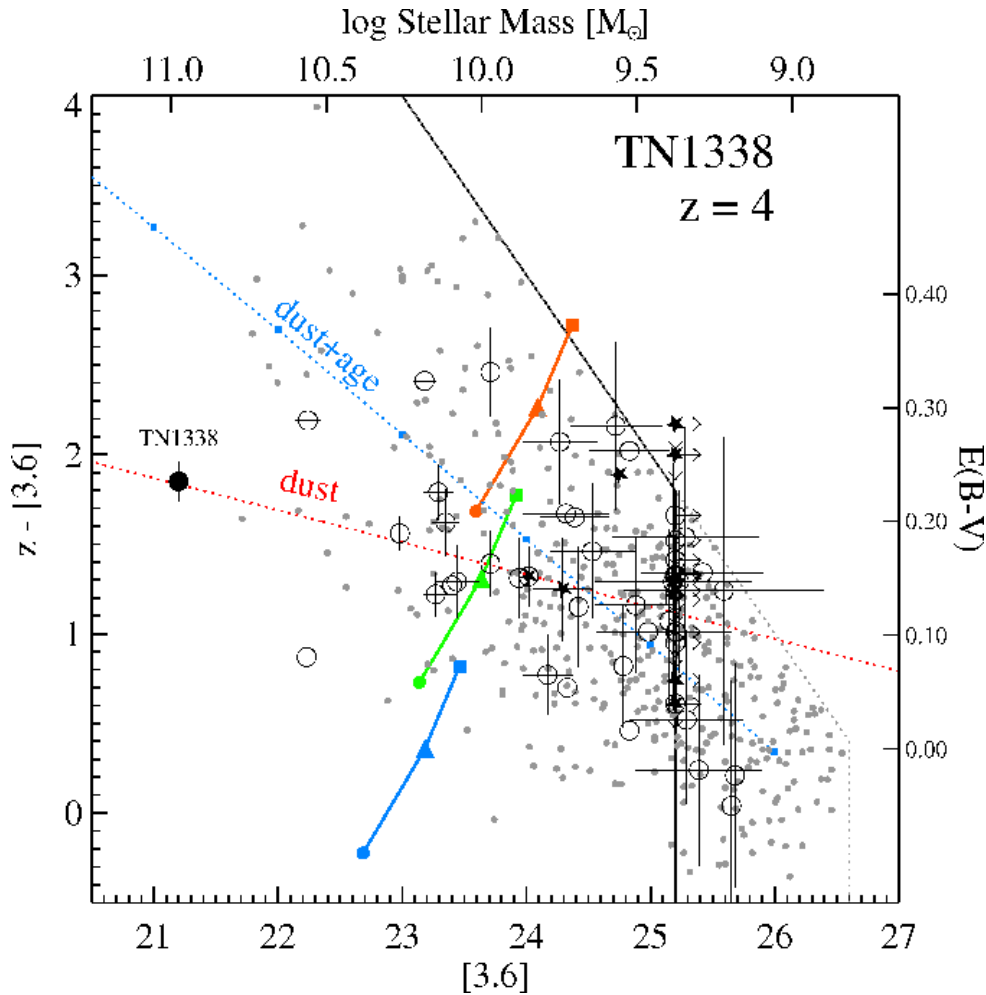} 
\end{center} 
\caption{The 3.6$\mu$m versus $z$--[3.6] color-magnitude
    diagram for galaxies in the field TN1338.
    Symbols indicate the radio galaxy (large filled circle), LBGs
    (open circles) and LAEs (filled stars). The 3.6$\mu$m flux of the
    radio galaxy has been corrected for \ha\ emission line
    contamination (see \S\ref{sec:ha}). The \bp-dropouts from
    \citet{stark09} are shown for comparison (small grey circles). The
    approximate detection limits for the TN1338 field and the GOODS
    control field are indicated by the black solid and grey dotted
    lines, respectively. The colored solid lines indicate our BC03
    baseline SED models described in \S\ref{sec:baseline},
    i.e. exponentially declining SFHs ($\tau=100$ Myr) having
    attenuations of $E(B-V)=0.0$ (blue), 0.15 mag (green) and 0.3 mag
    (red), each plotted at ages of 100 (circle), 200 (triangle) and
    300 Myr (square). The models were normalized to a total stellar
    mass of $10^{10}$ M$_\odot$. The right and top axes, respectively,
    show the attenuation and stellar mass obtained for a model having
    a fixed age of 200 Myr (for $M_*$ we fix $E(B-V)$ to 0.15). The red dotted line (marked `dust')
    indicates the minimum color change as a function of 3.6$\mu$m
    magnitude that is expected purely due to dust based on the
    $M_{UV}$ vs. $E(B-V)$ relation found by \citet{bouwens09} (and
    further assuming $\tau=100$, $t=200$ Myr in order to predict the
    $z$--[3.6] color). The blue dotted line (marked `dust$+$age')
    indicates a toy model in which we let both the dust and age
    increase in steps of $\Delta E(B-V)=0.05$ and $\Delta t=50$ Myr
    with decreasing 3.6$\mu$m magnitude starting from $E(B-V)=0.0$ and
    $t=200$ Myr (each step is indicated by a small blue square).  The
    latter model (`dust$+$age') is at least qualitatively in better
    agreement with the relatively steep change in color as a function
    of the 3.6$\mu$m magnitude compared to the `dust only' model.}
\label{fig:z4}
\end{figure*}

\section{RESULTS} 
\label{sec:results}

In this Section, we present the 3.6$\mu$m versus $z$--[3.6], or
rest-frame UV-optical, color-magnitude diagrams (CMDs) of the various
galaxy populations detected in the RG and QSO fields. We derive the
stellar masses of LBGs, LAEs and radio galaxies, and perform a
comparison between the samples in overdense regions and in the (GOODS)
field.

\subsection{UV-optical Color-Magnitude Diagrams in RG Fields at $z\simeq4-5$}

\begin{figure}[t]
\begin{center}
\includegraphics[width=\columnwidth]{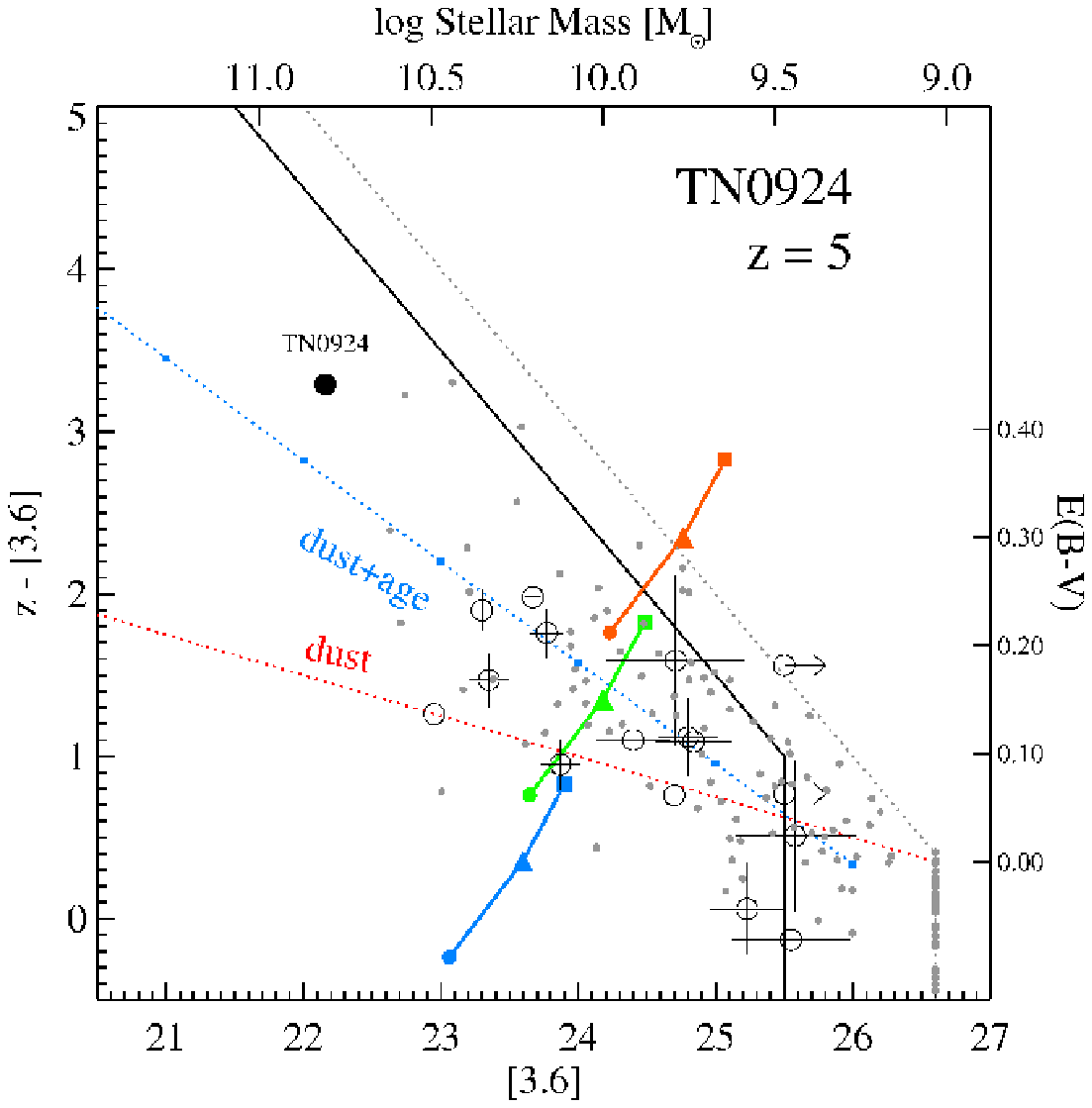}
\end{center}
\caption{The 3.6$\mu$m versus $z$--[3.6] color-magnitude diagram
  for galaxies in the field TN0924. Symbols
  indicate the radio galaxy (large filled circle) and LBGs (open
  circles).  The \vp-dropouts from \citet{stark09} are shown
  for comparison (small grey circles). The
    approximate detection limits for the TN0924 field and the GOODS
    control field are indicated by the black solid and grey dotted
    lines, respectively. The baseline SED model tracks are shown
  for $z\simeq5$ but are otherwise identical to the models in
  Fig. \ref{fig:z4} (see caption of Fig. \ref{fig:z4} for
  details). The right and top axes, respectively,
    show the attenuation and stellar mass obtained for a model having
    a fixed age of 200 Myr (for $M_*$ we fix $E(B-V)$ to 0.15). The red dotted line (marked `dust')
    indicates the minimum color change as a function of 3.6$\mu$m
    magnitude that is expected purely due to dust based on the
    $M_{UV}$ vs. $E(B-V)$ relation found by \citet{bouwens09} for
    $z\sim5$. The blue dotted line (marked `dust$+$age')
    indicates a toy model in which we let both the dust and age
    increase in steps of $\Delta E(B-V)=0.05$ and $\Delta t=50$ Myr
    with decreasing 3.6$\mu$m magnitude starting from $E(B-V)=0.0$ and
    $t=200$ Myr (each step is indicated by a small blue square).}
  \label{fig:z5}
\end{figure}

In Fig. \ref{fig:z4} we show the 3.6$\mu$m versus $z$--[3.6] 
diagram for our samples of galaxies at $z\simeq4.1$ in the field
TN1338. Symbols indicate the \gdrops (open circles), the LAEs (stars),
and the radio galaxy (large filled circle).  Two LAEs were
unfortunately confused beyond the possibility of deblending. Of the
remaining 10 LAEs, only three were detected at 3.6$\mu$m and we derive
limits for the seven that were undetected.  Also in Fig. \ref{fig:z4},
we show the distribution of $z\sim4$ \bp-band ($B$) dropouts from
GOODS (grey circles) as selected by \citet[][]{stark09} for
comparison.

Next, in Fig. \ref{fig:z5} we present the CMD of the \vdrops (open
circles) and the radio galaxy (filled circle) at $z\simeq5.2$ in the
field TN0924. Unfortunately, three out of four of the LAEs in this
field suffered from a large amount of confusion, and no attempt was
made for photometry. One LAE was found to be relatively isolated but
not detected. Again, for comparison with the field we indicate a large
number of \vdrops from GOODS using grey circles.

A number of basic observations from these diagrams can be made:\\

\noindent
(1) Our dropout samples show a sequence in the CMD with a total width
in color of $\sim2$ magnitude and a total range in 3.6$\mu$m of four
magnitudes ($\sim$22--26 mag), peaking at $3.6\mu m\sim25$ and
$z$--[3.6]$\sim1$. While the upper envelope in both CMDs is largely
set by the limiting $z$-band magnitude by which the samples were
selected (dashed lines in Figs. \ref{fig:z4} and \ref{fig:z5}), the
lower envelope is physical in the sense that brighter objects (at
3.6$\mu$m) are also redder. The same trend is seen for the GOODS
reference samples (grey circles), indicating that it is a general
feature of the high redshift population.\\

\noindent
(2) The LAEs in the TN1338 field, the only field with a sufficiently
large sample available for attempting photometry by means of
deblending (see \S2), mostly show non-detections at 3.6$\mu$m. The
three detected LAEs were L9, L16 and L22 \citep[see][]{overzier08}. L9
is also part of the LBG sample, while L16 and L22 are respectively too
blue or too faint to make it into our \gdrop LBG sample. These results
were expected given that about half of the LAEs are fainter than
\zp=26.5 magnitude in the rest-frame UV and the rather linear
correlation between the rest-UV and optical in the CMD. In any case we
can conclude that we have not found any LAEs that are particularly
bright or red compared to the LBGs, although we should note that the
RG has an extremely luminous associated \lya\ halo typical of radio
galaxies that would formally put it into a \lya-selected sample as
well.\\

\noindent
(3) Both radio galaxies have extraordinarily bright rest-frame optical
magnitudes, at least one magnitude brighter than the brightest LBG and
2--4 magnitudes brighter at 3.6$\mu$m compared to the bulk of the LBG
populations at $z\sim4$ and $z\sim5$. They also have relatively red
UV-optical colors but perhaps not redder than the extrapolated color
of the sequence formed by LBGs at lower magnitudes.

\subsection{Mass Segregation of LAEs, LBGs and Radio Galaxies}

As explained in \S\ref{sec:baseline} we use a limited set of spectral
energy distributions as simple baseline models for interpreting the
CMDs shown in Figs. \ref{fig:z4} and \ref{fig:z5}. In these figures we
have indicated the expected location of star-forming galaxies with
exponentially declining star formation histories of $\tau\approx100$
Myr, dust attenuations in terms of $E(B-V)$ of 0.0 (blue lines), 0.15
(green lines) and 0.3 mag (red lines) modeled using the recipes for
starburst galaxies given in \citet{calzetti01}, and plotted at three
ages of 100, 200 and 300 Myr (indicated by the circle, triangle and
square, respectively). The tracks were normalized to a total stellar
mass of $10^{10}$ M$_\odot$.  These models span an appropriate range
of parameter space compared to the detailed investigations of
\citet[e.g.][and references therein]{stark07,stark09} and \citet{bouwens09}.
The right and top axes of Figs. \ref{fig:z4} and \ref{fig:z5} 
show the attenuation and stellar mass obtained for a model having
a fixed age of 200 Myr.
 
\subsubsection{RGs} 

At $z=4.1$ and $z=5.2$, the 3.6$\mu$m band probes the rest-optical at
$\sim$7000 and 6000\AA, respectively. At these wavelengths, the light
is believed to be largely stellar in origin
\citep[][]{zirm03,seymour07}, despite the many significant,
non-stellar components of radio galaxies across the entire spectrum.
In both fields, we find that the RG is the most massive object, with
an inferred mass of $\sim$ $10^{11}$ \Msun. Without more constraints
on the SED we judge that this estimate can be higher or lower by
$\sim$25\% depending on the exact age and attenuation. TN0924 at
$z=5.2$ is over a magnitude redder than TN1338 at $z=4.1$. If the
attenuations are similar, it implies that the higher redshift RG is
about twice as old as the lower one ($\sim$600 vs. 300 Myr). In any
case, it is interesting that both RGs are much more massive than any
of the LBGs in their surrounding fields.

\subsubsection{LBGs} 

We now compare the dropouts with the model tracks plotted in
Figs. \ref{fig:z4} and \ref{fig:z5}.  In both RG fields we find that
the range in 3.6$\mu$m magnitude of the \gdrop and \vdrop LBGs implies
a distribution in stellar mass of about one order of magnitude, from a
few times $10^9$ to a few times $10^{10}$ M$_\odot$. The CMD slope can
be explained entirely by a correlation between dust and stellar mass,
finding ages of $\sim100-200$ Myr and dust increasing from $\sim0.0$
to $\sim$0.15 mag between the low and high mass end.  Alternatively,
interpreting the CMD slope as an age effect, the low mass end objects
are consistent with an age of $\sim$70--200 Myr and small attenuation,
while the typical age at the high mass end would be $\sim2.5\times$
higher ($\sim200-500$ Myr) for a similar attenuation. We will discuss
the age-dust (and metallicity) degeneracy in more detail in
Sect. \ref{sec:dust} below.

\subsubsection{LAEs} 

Due to the non-detections and overall small number of LAEs in the
TN0924 field, we can only compare LAEs and LBGs in the TN1338 field.
Most of LAEs were undetected, suggesting masses of $\lesssim10^9$
\Msun, and we note that many of the faintest LBGs were undetected as
well. The brighter, detected LAEs have colors that are consistent with
those of non-\lya\ LBGs at similar optical magnitudes.  The tendency
of LAEs at $z\sim4$ to be relatively faint and blue in the rest-frame
UV/optical is consistent with their young age as found in other
surveys
\citep[e.g.][]{shapley03,gawiser06,pentericci07,pentericci09,ouchi08}
and predicted by some simulations \citep{nagamine08}.

\subsection{Interpreting the relation between $M_*$ and $z$--[3.6] color}
\label{sec:dust}

In Figs. \ref{fig:z4} and \ref{fig:z5} we have observed a strong
  trend of redder $z$--[3.6] colors with increasing stellar
  mass. This trend is qualitatively similar to the ``blue envelope''
  of LBGs at $z\sim3-4$ observed by \citet{papovich04} and must be
  related to an increase in the mean age, dust or metallicity (or a
  combination thereof) with increasing rest-frame optical
  luminosity. While we cannot solve these degeneracies on an
  object-by-object basis without deep multi-wavelength photometry, we
  can use recent results from literature to explain the general trends
  observed. Deep studies of dropouts covering a wide range in both
  redshift and UV luminosity have now clearly demonstrated that
  star-forming galaxies become bluer in the UV continuum with increasing redshift and with
  decreasing luminosity
  \citep{papovich04,stanway05,hathi08,bouwens09}, in some contrast
  with earlier results that indicated no clear correlation between the
  slope of the UV continuum and UV luminosity
  \citep[e.g.][]{meurer99,adelberger00,ouchi04}. As shown by
  \citet{bouwens09}, these results can be fully explained by the
  relatively limited range in redshift and/or UV luminosity probed by
  earlier surveys. 
Using the deepest data available from the Hubble Ultra Deep Field
\citep[UDF;][]{beckwith06}, these authors show that there is a
significant correlation between the UV slope and the UV luminosity for
both $B$- and $V$-dropouts. They also investigate the relative effects
that, in principle, metallicity, age and dust each could have on the
observed UV slope. They find that both dust and age have the potential
of reddening the UV continuum significantly, with the effect of dust
being the strongest for realistic distributions of age and dust. The
influence of metallicity is found to be only small.

Could the trends that we found in Figs. \ref{fig:z4} and \ref{fig:z5}
be explained due to systematic changes in dust and age with increasing
stellar mass (or 3.6$\mu$m luminosity)? \citet{stark09} have shown
that the median ages of dropouts at $4<z<6$ (the same samples that we
have used for our comparison) change only little with either redshift
or UV luminosity (less than a factor of 2).  However, they do find a
strong correlation between stellar mass and SFR as inferred from the
UV continuum luminosity (with a net {\it specific} SFR, SFR/$M_*$,
that is relatively constant as a function of stellar mass). The
correlation between SFR and extinction by dust is well-known, as
galaxies with higher SFRs tend to have more dust
\citep[e.g.][]{wang96,martin05,reddy06}. \citet{bouwens09} have now
firmly established this correlation for LBGs by studying the average
UV color (a measure of reddening) as a function of the absolute UV
magnitude (a measure of the unextincted SFR). As argued by
\citet{bouwens09}, we thus expect to see a strong correlation between
stellar mass and reddening due to dust. In Figs. \ref{fig:z4} and
\ref{fig:z5} we have indicated the expected amount of reddening purely
due to dust based on the empirical UDF results. In both figures, the red
dotted line marked `dust' indicates the color change (as a function of
3.6$\mu$m magnitude) expected based on the relation between $M_{UV}$
vs. $E(B-V)$ from \citet[][see their Fig. 3]{bouwens09}. Using this
relation, the extinction varies from $E(B-V)\approx$0.0 to $\approx$0.2 mag from the
faint to the bright end. While the trend predicted by our UV-to-($z$--[3.6]) color
extrapolation shows the correct general behavior, the reddening in
$z$--[3.6] expected seems to underpredict the amount of reddening
observed. This may indicate that, besides the increased reddening due
to dust, additional reddening is required due to age as a function of
stellar mass. 

To illustrate this further we construct a simple toy model. We let
both dust and age increase towards brighter 3.6$\mu$m magnitudes. 
The $E(B-V)$ is increased in steps of 0.05 and the age in
steps of 50 Myr. The resulting color trend for this toy model is shown
by the blue dotted line marked `dust$+$age' in Figs. \ref{fig:z4} and \ref{fig:z5}. The
starting point (marked by the small blue square on the faint end of
the blue dotted line) was chosen to have $E(B-V)=0.0$ and an age of 200 Myr.
The `dust$+$age' model is, at least qualitatively, in better
agreement with the relatively steep change in color as a function of
the 3.6$\mu$m magnitude compared to the `dust'-only model. 

In principle, it should be possible to find the required evolution in
dust and age that best fits the general trends observed in the
CMDs. However, Figs. \ref{fig:z4} and \ref{fig:z5} show that the
dynamic range in $z$--[3.6] color in the TN1338, TN0924
and GOODS fields is too small to carry out such an exercise (i.e., a
fit to the data can not be made due to relatively red sources being missed at the faint end). Deeper
data in the rest-frame UV would greatly alleviate this problem, and
it should be possible to perform such a study in the much deeper UDF. 

\subsection{Comparison of galaxies in overdense regions versus galaxies in average environments from GOODS}

It has previously been shown that the {\it number density} of dropouts
and LAEs in the TN1338 and TN0924 fields is larger relative to random
fields, indicating that these fields may host massive galaxy
clusters-in-formation
\citep{venemans02,venemans04,overzier06,overzier08}. One of the goals
of the current paper was to test whether besides this enhancement in
number density there also exist differences in the physical properties
of galaxies in overdense regions compared to the field. The large
samples of $B$- and $V$-dropouts extracted from the $\sim320$
arcmin$^2$ GOODS field by \citet{stark09} allow us to make such a
comparison. Comparing the color-magnitude distribution in our fields
with the results found for GOODS (grey circles in Figs. \ref{fig:z4}
and \ref{fig:z5}) indicates a very similar trend of redder colors for
brighter objects. This suggests that, at least at first sight, in both
types of environment the dropout populations as a whole possess very
similar stellar populations.

Unfortunately, there are several factors that complicate making
further quantitative statements about the differences or similarities
between overdense fields and GOODS. First, the size of the individual
error bars in the CMD diagrams is substantial, thereby limiting the
usefulness of object-by-object comparisons except for the brightest
objects (as we will see below). Second, because we only have limited
filters and miss deep NIR data, an analysis based on multi-band SED
fitting is not yet possible for our fields. Third, although many of
the dropout galaxies are believed to be associated with the
protoclusters in the vicinity of the radio galaxies, we must note that
the dropouts in TN1338/TN0924 are not spectroscopically confirmed,
except for the subset that are \lya-bright. This means that our
dropout samples will undoubtedly include galaxies in the (near) fore-
and background of the protoclusters that could offset or bias any
trends observed in Figs. \ref{fig:z4} and \ref{fig:z5}.
 
These caveats aside, however, we can at least ask the question what
differences with GOODS we would in principle be able to detect based
on the quality of our data presented. The first possibility we
consider is that if galaxies in the overdense regions were
systematically bluer or redder compared to the field, the protocluster
dropouts may be expected to form a lower or upper envelope with
respect to the well-established CMD slope formed by the GOODS dropouts
at $z\simeq4-5$. We estimate that a systematic difference in ages by a
factor of $\gtrsim2$ or a difference in attenuation of $\gtrsim0.15$
mag would have been detectable given the size of the error bars in
Fig. \ref{fig:z4}. However, no such differences are suggested by the
data.

Another possibility we consider concerns the hypothesis that the
overdense regions contain more massive galaxies compared to the
general field. In this scenario, we would expect an extension towards
brighter 3.6$\mu$m magnitudes (larger stellar masses) relative to
GOODS. Because every object in the overdense region would become a
little brighter, this means that at the faint end of the
luminosity/mass function more objects in the overdense region would
pass the detection threshold, while the number for the ``field'' would
remain the same. Therefore, one would not immediately expect to find
any differences between the average properties of galaxies at the
faint end. At the bright end, however, one would expect an excess of
massive sources relative to the field. We estimate that a boost in
masses by a factor of a few should be detectable in the CMD
diagrams. In this respect, it is interesting that in both fields, the
RGs are brighter than any dropout from the GOODS comparison sample,
even though the latter survey covers an area that is over 25 times
larger. We have also analyzed the relative numbers of 
dropouts in the fainter magnitude bins, finding no 
evidence for a significant brightening of dropouts in the overdense
regions. From this it follows that compared to GOODS the protocluster
fields likely do not contain an excessive number of bright/massive
($\sim10^{11}$ $M_\odot$) LBGs other than the radio galaxies
themselves. We will further discuss these results in \S\ref{sec:discussion}.

\begin{figure}[t]
\begin{center}
\includegraphics[width=\columnwidth]{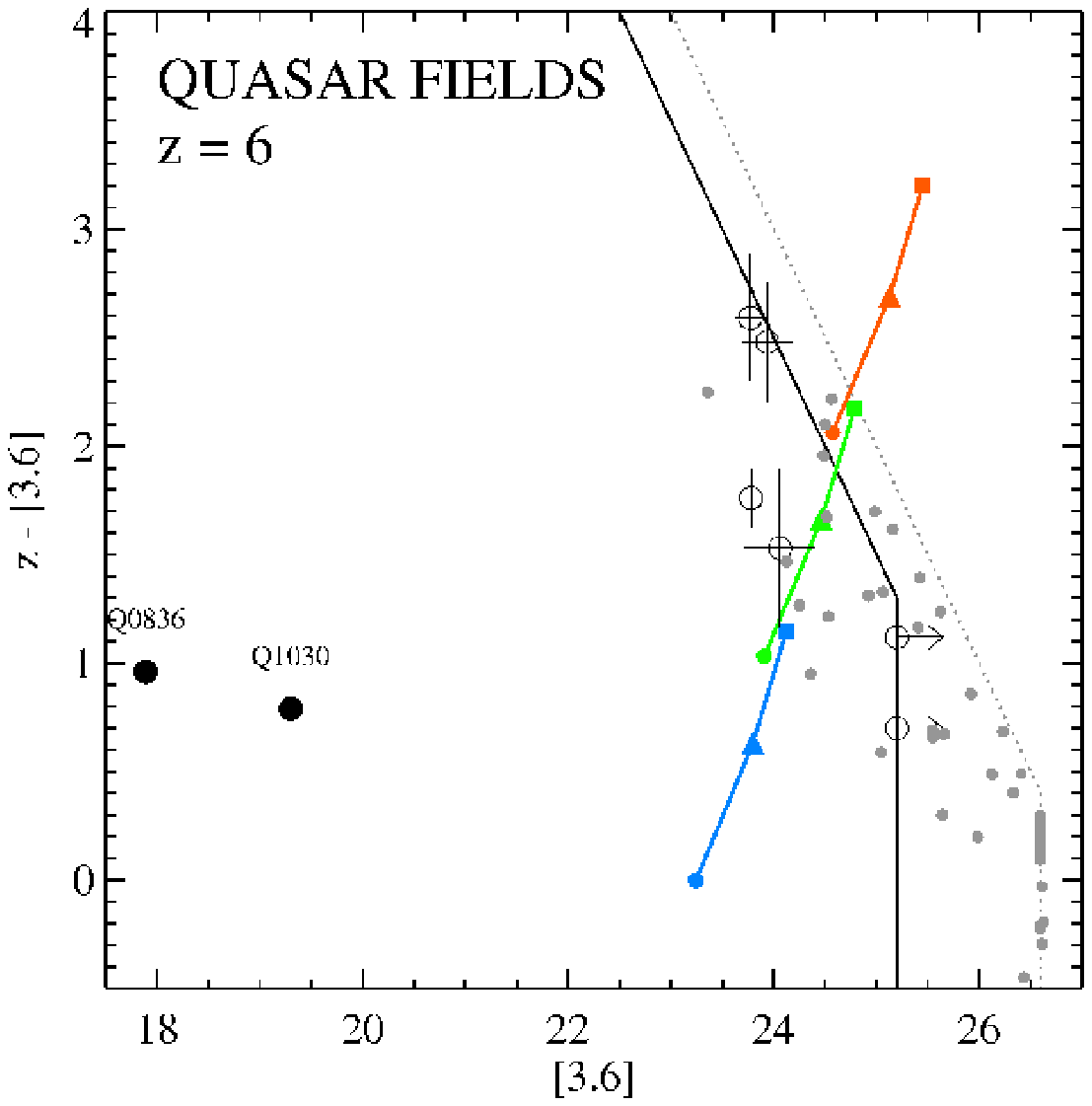} 
\end{center}
\caption{3.6$\mu$m versus $z$--[3.6] color-magnitude diagram for
  the $z\simeq6$ \ip-dropouts selected from the fields SDSS0836 (QSO
  at $z=5.82$) and SDSS1030 (QSO at $z=6.28$). Symbols indicate the
  quasars (large filled circles) and \ip-dropouts (open circles).  The
  \ip-dropouts from \citet{stark09} are shown for comparison
  (small grey circles). The
    approximate detection limits for the QSO fields and the GOODS
    control field are indicated by the black solid and grey dotted
    lines, respectively. 
Model tracks have been indicated for
  $z\simeq6$ but are otherwise identical to the ones shown in
  Figs. \ref{fig:z4} and Fig. \ref{fig:z5} (see the caption of
  Fig. \ref{fig:z4} for details). The QSOs have been indicated for
  reference only, as their magnitude and colors are completely
  dominated by AGN light instead of stellar light.}
\label{fig:z6} 
\end{figure}

\subsection{UV-optical Photometry of $i$-dropouts in Two QSO Fields at $z\simeq6$} 

For completeness, we will use our ACS and IRAC data on the two
$z\sim6$ QSO fields in order to perform a similar analysis for the
\idrop candidates in the (projected) vicinity of the luminous
quasars. We have tabulated our measurements in Table 4 and plot the
CMD in Fig. \ref{fig:z6}. We also plot the QSOs in the CMD but stress
that their emission is strictly dominated by non-stellar processes and
should not be compared with any other populations or models shown in
this paper.  We compare our results again with a small sample of
similarly selected \idrops from GOODS (grey circles), and with the
same model tracks as plotted in Figs. \ref{fig:z4} and
\ref{fig:z5}. The colors of the four \idrops in the QSO fields that
are detected in both bands are clearly inconsistent with very young,
dust-free models (e.g. the blue track in Fig. \ref{fig:z6}), while the
masses are on the order of $10^{10}$ \Msun\ relatively independent of
dust and age.  The faintest objects that can be probed in GOODS are
about ten times less massive due to the much higher sensitivity in the
[3.6] channel \citep{yan05,stark07}. Comparing the two
distributions further suggests an excess of sources having
3.6$\mu$m$<$24.5 magnitude in the QSO fields not seen in
GOODS. Unfortunately, the small number statistics, the uncertainties
in the redshifts and the photometric errors do not allow us to say at
present whether this trend is real or not.
 
\section{SUMMARY AND DISCUSSION}
\label{sec:discussion}
 
\subsection{Summary}

Summarizing our results from the previous section, at $z\simeq4.1$,
the masses and stellar populations of RGs and LBGs are found to lie in a
similar range as those at $z\simeq 5.2$, with the RGs being
systematically the most massive objects ($\sim10^{11}$
$M_\odot$). The stellar mass of the LBGs ranges from a few times $10^9$ to a few times
$10^{10}$ M$_\odot$, and there is a clear trend of redder colors with
increasing mass. The correlation can be explained if more massive
galaxies have more dust (because their SFR is higher) and are somewhat 
older compared to less massive galaxies. 
Few LAEs were detected, implying they are generally
of low mass. The RGs, LBGs and perhaps LAEs thus appear to be on
different evolutionary tracks with the RGs having experienced either a
more prolonged or more efficient phase of star formation history
compared to LBGs, and LAEs being on average in earlier stages than the
typical UV-selected LBGs in these fields (but perhaps not different
compared to LBGs of the same UV luminosity!). We investigated possible
differences between the properties of galaxies in overdense regions
and those of dropouts from the much larger GOODS survey based on the
distribution of objects in the rest-frame UV-optical CMD diagram. We
found no significant, systematic differences between the overall
distributions in $z$--[3.6] color and 3.6$\mu$m magnitude between
dropout galaxies in the different regions.  We did however find that
in both fields the stellar masses of the radio galaxies as inferred
from the 3.6$\mu$m flux are highly excessive compared to the general
field expectations from GOODS. The latter survey virtually contains no
dropout galaxies as bright (at 3.6$\mu$m) as the RGs, even though the
area surveyed by GOODS is about 25$\times$ larger. Their (inferred)
stellar masses of $\sim10^{11}$M$_\odot$ are also about an order of
magnitude larger than the brightest LBGs in the protocluster fields.

\subsection{The most massive galaxies at $z\simeq4-5$}

Even though we have shown that none of the objects in GOODS are as
bright as the RGs, \citet{stark09} find a non-negligible number of
objects having comparable masses based on the detailed SED fits
performed for GOODS. Thus, for the first time, we have a quantitative
number density of massive galaxies from the field to be compared with
RGs and protocluster regions.  In order to quantify how rare galaxies
with stellar masses as high as observed for the RGs are, we use the
total number densities of (UV-selected) galaxies having stellar masses
$>10^{11}$ \Msun\ as derived by \citet{stark09}. They find number
densities of $N(>M)\sim6\times10^{-6}-4\times10^{-5}$ Mpc$^{-3}$ at
$z\sim4$ and $N(>M)\sim1\times10^{-6}-2\times10^{-5}$ Mpc$^{-3}$ at
$z\sim5$ (ranges are 1$\sigma$), and we note that the upper values on
these number densities are perhaps overestimated as
non-spectroscopically confirmed sources having
$z$--[3.6]$\gtrsim2$ may have some contamination from dusty
star-forming systems at $z\simeq2$ \citep{yan04,stark09}.

The co-moving volumes of the TN1338 and TN0924 fields are
$\approx2\times10^4$ Mpc$^3$ \citep{overzier06,overzier08}, implying
that we would naively expect as many as, respectively, 0.1--0.8 (0.1--0.6 if
we assume that the 24$\mu$m-detected objects in GOODS are foreground
contamination) and 0.02--0.4 highly massive galaxies in a random
single ACS pointing. Thus, it appears that the presence of a single
$\sim10^{11}$ \Msun\ (radio) galaxy in the protocluster fields is
marginally consistent with the ($1\sigma$) upper limits on the cosmic
average number density at $z\simeq4$ while it is a positive deviation
of several sigma at $z\simeq5$. The GOODS field estimates are about
two times higher than the values derived by \citet{mclure08} based on
a much larger survey of 0.6 deg$^2$, indicating that the GOODS
estimates are probably not missing a significant fraction of massive
(UV-selected) objects. Furthermore, to date no objects have been found
with masses in excess of a few times $10^{11}$ \Msun. Our results thus
confirm earlier studies that suggest that RGs are among the most massive galaxies in
the early Universe \citep[e.g.][and see Fig. 5 in \citet{rocca04}]{pentericci97,seymour07}. The low average number density of such objects
further implies that the RGs in the TN1338 and TN0924 fields have
experienced a rapid growth through accretion of gas or merging,
presumably as a consequence of their overdense environments. This is
qualitatively consistent with some models that specifically require an
extra bias factor due to mergers in order to explain the strong
clustering of luminous quasars at $z\sim4$ \citep[e.g.][]{wyithe09}.

From the lack of any known bright radio sources among the massive
GOODS galaxies \citep[see][for details on a radio follow-up in the
  GOODS-CDFS]{mainieri08}, we can deduce that {\it if} a powerful
(radio) AGN is a typical phase of massive galaxies at
$z\simeq4-5$, then it is unlikely to last longer than 30--50
Myr. Otherwise, radio galaxies would be quite common in a field the
size of GOODS given that the \bdrop ($V$-dropout) samples span
$\sim$500 ($\sim$300) Myr in look-back time. This simple estimate
based on the number counts is quite close to the expected radio source
life-times of $\sim$10 Myr derived using other arguments
\citep{blundell99}. If, on the other hand, none of the massive GOODS
galaxies become radio galaxies even for a brief period during the
epoch at $z\simeq4-5$, it may mean that additional requirements for
producing them must be met. Perhaps the overdense environment as
observed in the protocluster fields TN1338 and TN0924 facilitates the
rapid accretion of cold gas through mergers or cosmological flows
\citep{keres09} spawning both a massive starburst and powerful AGN
activity of a supermassive black hole \citep[see][]{zirm05,zirm09}.

\subsection{Relation to low redshift clusters}

\citet{stark09} compared the number density of $\gtrsim10^{11}$
M$_\odot$ $B-$ and \vdrop systems to the observed density of quiescent
$\sim10^{11}$ M$_\odot$ galaxies at $z\sim2.5$, finding that the high
redshift sample can account for $\sim20$\% of the massive population
at lower redshift.  In a similar fashion, we can compare the number
density of the most massive objects in GOODS at these redshifts to the
number density of galaxy clusters in the local Universe, finding that
they are roughly equal for clusters having $L_X\sim1-5\times10^{43}$
erg s$^{-1}$ \citep{rosati98}. Thus, one may expect that the $z=0$
descendants of the most massive $z\sim4$ galaxies in GOODS are found
in typical group and cluster environments \citep[see
  also][]{ouchi04}. Since the radio galaxy fields studied in this
paper are overdense with respect to GOODS, it will be interesting to
perform a systematic study of galaxies in the direct environment of
the massive GOODS galaxies \citep[e.g.][]{daddi08}.

We have argued that the overdense regions correspond to sites of
massive cluster formation. If so, we expect that the protocluster
region will start to develop a cluster red sequence and a virialized
intracluster medium at some redshift $z\lesssim2$. Likely, the stars
formed in the protocluster galaxies observed at $z\simeq4-5$ will
evolve and become incorporated into the population of massive
early-type galaxies on the red sequence. We can use our results of
\S\ref{sec:results} to estimate what fraction of this low redshift
cluster red sequence mass is represented by all protocluster galaxies
detected. We use the conversion between 3.6$\mu$m magnitude and $M_*$
derived from the simulations in \S\ref{sec:mrtest} and calculate the
total mass of all dropouts in the TN1338 field at $z=4$. We find a
total stellar mass of $\sim5\times10^{11}$ $M_\odot$. About 25\% of
this mass is due to the large contribution from the radio galaxy.

We will now compare our estimate to the total stellar mass on the
  red sequence of the cluster RDCS1252.9--2927 (Cl1252), a massive
  X-ray luminous cluster at $z=1.24$
  \citep{blakeslee03a,rosati04,demarco07}. Although we do not
  necessarily expect that TN1338 will evolve into a similar cluster as
  Cl1252 (see discussion below), we choose this particular cluster for
  our demonstration for a number of reasons. First, the analysis of
  Cl1252 was performed by members of our team, giving a good
  understanding of the data and its interpretation. More importantly,
  it is one of the best studied high redshift clusters, and its high
  redshift ensures a minimal gap in cosmic time with respect to $z=4$
  (about 3.5 Gyr). This is important because below we will perform a
  simple extrapolation of the stellar masses measured at $z\sim1$ to
  those predicted at $z=4$.

The analysis of Cl1252 performed by \citet{rettura08} found that 80\%
of the ETGs have population ages of $3.5\pm1$ Gyr, consistent with a
mean formation redshift of $z\sim4$. If we limit ourselves to ETGs for
which the mean star formation weighted ages are $>3.5$ Gyr, we find 10
massive ($M_*\gtrsim10^{11}$ $M_\odot$) early-type galaxies (ETGs)
representing a total ``red sequence mass'' of $3.5\times10^{12}$
$M_\odot$ (about half of which is due to the three most massive ETGs).
If we now assume that all the dropouts detected in the TN1338 field
are part of the protocluster, its total stellar mass amounts to about
14\% (at most) of the total Cl1252 ETG mass presumed to have formed at
$z\gtrsim4$. This percentage (14\%) implies that we are currently
missing most of the mass that we naively expected to have existed
already at $z\sim4$. Where could this mass be hiding? We suggest four
possible explanations for this apparent discrepancy:\\

(1) When evolved to $z\sim1$, TN1338 could be a very different cluster
than Cl1252. On one hand, large uncertainties in the physical sizes
and the magnitude of the (dark matter) overdensities make it currently
very difficult to predict the total mass and virialization redshift of
the descendants of protoclusters. Estimates indicate that their total
masses range from $10^{14}$ to $10^{15}$ $M_\odot$, with virialization
redshifts of $0\lesssim z\lesssim1$
\citep[e.g.][]{steidel05,overzier08}. On the other hand, Cl1252 is an
exceptionally massive, X-ray luminous cluster indicating virialization
that was largely complete at $z=1.24$. The relatively low total
stellar mass found for TN1338 could therefore indicate that it will
evolve into a more typical and lower mass cluster, containing fewer
ETGs on the red sequence at $z\lesssim1$ compared to
Cl1252. Interestingly, the existence of extremely massive clusters
such as Cl1252 at $z\sim1.2$ may indicate that much larger
protoclusters at $z>2$ remain to be discovered.\\

(2) A large fraction of the ETG mass may be accreted from a much larger region that extends beyond the region probed by our ACS pointing. The \lya\ emitters found by \citet{venemans07} extend significantly beyond the area covered by our HST data, and \citet{intema06}  found evidence that TN1338 is part of a larger structure of LBGs at a co-moving scale of $\sim$10 Mpc.\\

(3) The stars may be contained in numerous sub-units that are too
faint to be detected by our current survey. Extending the luminosity
function of $z\sim4$ $B$-dropouts down to significantly below the
detection limit suggests that the contribution from faint objects to
the total star formation rate amounts to $\lesssim$50\%
\citep[e.g.][]{bouwens07}. We may therefore increase our total mass
estimate in the overdense region by no more than about a factor of 2,
increasing the ETG progenitor mass accounted for at $z\sim4$ to
$\sim$30\%. If this very faint high-redshift population provides
indeed an important contribution to the mass of ETGs, they will need
to merge to form the ETG population within a few Gyr.\\

(4) A significant fraction of the stellar mass may be contained in
systems that are either very dusty or have little on-going star
formation and are hence missed by our UV-selection.\\

The answers to questions (1) and (2) are currently difficult to
address based on the limited observational data and the small number
of protoclusters known. These issues will be the subject of a future
paper in a series of papers aimed at comparing the observational
properties of statistical samples of (proto-)clusters extracted from
cosmological simulations of cluster formation \citep[][and in
  prep.]{overzier09}. The possible incompleteness due to red galaxies
(question 4) will be discussed in the next section.

\subsection{Have we missed a significant population of quiescent or obscured galaxies?}
\label{sec:raredust}

Because UV-selected samples may miss galaxies that are either
quiescent or obscured, we may expect that the estimates for the number
density of massive galaxies both in GOODS and in the protocluster
fields provide a lower limit to the actual number densities. At
$z\gtrsim2$ there is a substantial contribution to the stellar mass
function from the populations of so-called distant red galaxies
(DRGs), which consist of both obscured, star-forming objects and red,
quiescent objects that tend to be substantially more massive (and
older) compared to UV-selected LBGs
\citep[e.g.][]{franx03,brammer07,dunlop07,wiklind07}. If such objects are abundant in
protocluster regions, we may largely resolve the discrepancy we found
above between the total stellar mass observed and that expected based
on $z\sim1$ ETGs in Cl1252. Will the overdensities in LBGs and LAEs
observed in TN1338 and TN0924 be accompanied by similar overdensities
in relatively red or massive galaxies? \citet{brammer07} have shown
that the UV-selection at $z\sim2-3$ may miss as much as $\sim33$\% of
galaxies having very red UV continuum slopes that are selected on the
basis of a strong Balmer Break. However, at a higher redshift of
$z\sim3-4.5$ almost all ($\gtrsim90$\%) galaxies in their Balmer Break
sample are also selected by means of a UV selection. \citet{bouwens09}
show that the UV-dropout selections in GOODS and the UDF at
$z\gtrsim2$ become substantially more complete with increasing
redshift. This is because the UV-continuum slope of the galaxy
population becomes bluer as redshift increases. At $z\gtrsim5$ the
distribution of UV slopes of the entire dropout population lies within
the dropout selection function, and even at $z\sim4$ the selection is
believed to be largely complete consistent with, e.g., \citet{brammer07} and \citet{dunlop07}.
Finally, we consider the population of ultra-luminous IR galaxies
(ULIRGs). The high redshift ULIRG population is a population of
heavily obscured starbursts detected in the sub-mm and corresponds to
rapidly forming, massive galaxies at $z\simeq2-4$. However, analogous
to the population of Balmer break objects, the contribution of ULIRGs
to the star formation rate density declines with redshift from
$\sim$30\% at $z\sim2.5$ to $\sim10$\% at $z\sim4$
\citep{capak08,daddi08,bouwens09}. Therefore, we currently do not have
very strong reasons for believing that a large population of either
the red Balmer break objects or heavily enshrouded starbursts is being
missed in our TN1338 protocluster at $z=4$.  Follow-up studies are currently being undertaken (C. De Breuck, private communication) based on
deep near-infrared data specifically targeting galaxies having
prominent Balmer breaks between the $H$ and $K$ bands ($K$ and
3.6$\mu$m) for $z\sim4$ ($z\sim5-6$) that could resolve whether
protoclusters have an overabundance of such red objects compared to
the general field. 


\acknowledgments 
We thank Chien Peng for help with GALFIT. This research has been supported by NASA through Grant GO20749 issued by 
JPL/Caltech and Grant NAG5-7697 to the Johns Hopkins University.

\begin{deluxetable}{llllll}
\tabletypesize{\scriptsize}
\tablecolumns{6}
\tablewidth{0pc}
\tablecaption{\label{tab:samples}Overview of Targeted Fields}
\tablehead{\multicolumn{1}{c}{ID} & Redshift & \multicolumn{1}{c}{$\alpha_{J2000}$} & \multicolumn{1}{c}{$\delta_{J2000}$} & Type & Galaxy Samples}
\startdata
TN J1338--1942   & 4.11 &  13:38:26.06 & --19:42:30.5 & Radio Galaxy           & $g$-dropouts, \lya\ emitters\\
TN J0924--2201   & 5.20 &  09:24:19.89 & --22:01:41.3 & Radio Galaxy           & $V$-dropouts, \lya\ emitters\\
SDSS J0836+0054 & 5.82 &  08:36:43.87 &     $+$00:54:53.2 & Radio-Loud Quasar & $i$-dropouts\\
SDSS J1030+0524 & 6.28 &  10:30:27.09 &     $+$05:24:55.0 & Radio-Quiet Quasar & $i$-dropouts\\
\enddata
\end{deluxetable}

\begin{deluxetable}{lllrrrrcrr}
\tabletypesize\small
\tablecolumns{9}
\tablewidth{0pc}
\tablecaption{\label{tab:lbgz4}Photometric Properties of $g_{475}$-Dropouts 
in the Field of TN J1338--1942.}
\tablehead{\multicolumn{1}{c}{ID$^a$} & \multicolumn{1}{c}{$\alpha_{J2000}$} 
& \multicolumn{1}{c}{$\delta_{J2000}$} &\multicolumn{1}{c}{$i_{\rm 775} - z_{\rm 850}$}
 & \multicolumn{1}{c}{$z_{850}^b$} & \multicolumn{1}{c}{$3.6 \mu m^c$} 
& \multicolumn{1}{c}{$4.5 \mu m^c$} & \multicolumn{1}{c}{Class$^d$} }
\startdata
  $  2707/\rm RG^f$       &13:38:26.06 & --19:42:30.50&  $0.09\pm 0.03$   
&$23.05\pm0.05$& $20.45\pm0.03$ & $21.10\pm0.02$ & 3 \\
  $  367$       &13:38:32.74 & --19:44:37.29 & $0.07\pm 0.02$ 
&$23.10\pm0.02$& $22.23\pm0.08$ & $22.35\pm0.10$ & 3 \\
  $  1991$       &13:38:27.83 & --19:43:15.36 &$0.05\pm 0.07$ 
&$24.43\pm0.12$& $22.24\pm0.10$ &$22.09\pm0.09$ & 3 \\
  $  3018$       &13:38:24.30 & --19:42:58.13 &$-0.01\pm 0.05$ 
&$24.49\pm0.07$& $23.27\pm0.10$ & $23.39\pm0.13$ & 3 \\
  $  3216$       &13:38:22.37 & --19:43:32.44 &$0.10\pm 0.05$
 &$24.54\pm0.06$& $22.98\pm0.07$ & $22.79\pm0.07$& 3 \\
    $  3116$       &13:38:24.21 & --19:42:41.88 &$-0.00\pm 0.06$ 
&$24.67\pm0.06$& $23.40\pm0.10$ &$23.00\pm0.09$ & 3 \\
    $  959$       &13:38:32.66 & --19:43:03.70  &$0.28\pm 0.07$
&$24.73\pm0.09$& $23.44\pm0.18$ & $23.36\pm0.17$& 3 \\
    $  2913$       &13:38:23.67 & --19:43:36.66 &$-0.01\pm 0.06 $ 
&$24.94\pm0.10$& $24.17\pm0.20$ &$24.11\pm0.26$ & 3 \\
    $  2152$       &13:38:26.91 & --19:43:27.60 &$ -0.01\pm 0.08$ 
&$24.97\pm0.15$& $23.35\pm0.11$ &$23.87\pm0.31$ & 3 \\
$  2799$       &13:38:24.88 & --19:43:07.40 &$-0.03\pm 0.07$ 
&$25.03\pm0.09$& $24.33\pm0.25$ &$>25.0$& 1 \\
$  2439$       &13:38:25.34 & --19:43:43.67 &$0.14\pm 0.08$  
&$25.08\pm0.10$& $23.29\pm0.12$ &$23.30\pm0.12$& 3 \\
$  3430$       &13:38:21.21 & --19:43:41.98 &$-0.03\pm 0.07$ 
&$25.10\pm0.09$& $23.71\pm0.16$&$23.84\pm0.20$ & 1 \\
$  2407$       &13:38:24.35 & --19:44:29.15 &$0.04\pm 0.08$
 &$25.11\pm0.11$& $\dots$&$\dots$ & 4 \\
$  2839$       &13:38:25.90 & --19:42:18.39 &$0.12\pm 0.07$ 
&$25.14\pm0.10$& $\dots$&\dots & 4 \\
$  1252$       &13:38:31.98 & --19:42:37.47 &$-0.03\pm 0.06 $
 &$25.25\pm0.08$& $23.94\pm0.21$&$24.01\pm0.26$ & 3 \\
$  227$        &13:38:33.02 & --19:44:47.53 &$-0.14\pm 0.09$ 
&$25.29\pm0.19$& $24.83\pm0.39$ &$24.40\pm0.38$& 3 \\
$  2710/L9^e$       &13:38:25.09 & --19:43:10.77 &$-0.02\pm 0.07$ 
&$25.34\pm0.08$& $24.02\pm0.15$&$24.37\pm0.34$ & 1 \\
$  1815$       &13:38:29.01 & --19:43:03.28&$0.01\pm 0.10$  
&$25.50\pm0.12$& $\dots$ & $\dots$ &4 \\
$  1152$       &13:38:32.63 & --19:42:25.17 &$-0.19\pm 0.11$
 &$25.57\pm0.20$& $24.42\pm0.27$ &$24.97\pm0.53$ &  3 \\
$  2755$       &13:38:24.92 & --19:43:16.93 &$0.12\pm 0.18$
 &$25.59\pm0.21$& $23.18\pm0.09$&$22.81\pm0.09$ &  1 \\
$  3304$       &13:38:23.67 & --19:42:27.39 &$ 0.08\pm 0.10$ 
&$25.60\pm0.12$& $24.78\pm0.32$ &$>25.0$ & 1 \\
$  1819$       &13:38:29.61 & --19:42:38.21 &$0.15\pm 0.09$ 
&$25.60\pm0.15$& $\dots$ & $\dots$ &4 \\
$  3159$       &13:38:22.21 & --19:43:50.13 &$0.11\pm 0.20$ 
&$25.63\pm0.16$& $25.39\pm0.51$ & $>25.0$ &1 \\
$  309$       &13:38:34.77 & --19:43:27.62  &$0.06\pm 0.12$
&$25.69\pm0.15$& $25.65\pm0.68$ &$>25.0$ &1 \\
$  1808$       &13:38:30.04 & --19:42:22.50 &$ -0.10\pm 0.10$ 
&$25.71\pm0.12$& $\dots$ & $\dots$ &4 \\
$  3670$       &13:38:20.73 & --19:43:16.34 &$0.07\pm 0.10$ 
&$25.81\pm0.11$& $25.29\pm0.46$ &$>25.0$ &1 \\
$  633/L25^e$       &13:38:34.96 & --19:42:24.98 &$ 0.00\pm 0.12$ 
&$25.81\pm0.15$& $>25.2$ &$>25.0$ &2 \\
$  2524$       &13:38:24.47 & --19:44:07.25  &$-0.02\pm 0.12$
&$25.86\pm0.21$& $\dots$ & $\dots$ & 4 \\
$  1461$       &13:38:31.37 & --19:42:30.94  &$0.07\pm 0.13 $ 
&$25.89\pm0.15$& $25.68\pm0.61$ & $24.77\pm0.34$& 3 \\
$  3177$       &13:38:22.97 & --19:43:16.08  &$0.12\pm 0.15$
&$25.99\pm0.23$& $24.98\pm0.42$ & $>25.0$ &1 \\
$  1668$       &13:38:26.94 & --19:44:53.24  &$0.16\pm 0.13$
&$25.99\pm0.16$& $24.53\pm0.34$ &$25.0\pm0.75$ &3 \\
$  358$       &13 38 32.13 & --19:45:04.67  &$0.03\pm 0.13 $
&$25.99\pm0.19$& $24.32\pm0.35$ &$24.51\pm0.45$ &3 \\
$  2569$       &13 38 26.38 & --19:42:43.53 &$0.03\pm 0.13 $ 
&$26.04\pm0.15$& $24.39\pm0.28$ &$25.16\pm0.85$ &3 \\
$  3131$       &13 38 25.27 & --19:41:55.54 &$-0.09\pm 0.14$
 &$26.04\pm0.19$& $24.88\pm0.33$ &$>25.0$ &3 \\
$  2527$       &13 38 27.99 & --19:41:44.08 &$-0.13\pm 0.10$
 &$26.11\pm0.22$& $\dots$ &$\dots$ &4 \\
$  2347$       &13 38 27.99 & --19:42:12.22 &$0.32\pm 0.16$
 &$26.14\pm0.19$& $\dots$ &$\dots$ &4 \\
$  2358$       &13 38 24.12 & --19:44:47.21 &$ -0.12\pm 0.15$
 &$26.15\pm0.16$& $>25.2$ &$>25.0$ &2 \\
$  307$       &13:38:32.83 & --19:44:46.28  &$-0.04\pm 0.13$
&$26.17\pm0.22$& $23.71\pm0.12$ & $23.09\pm0.10$ &3 \\
$  2989$       &13:38:22.90 & --19:43:59.01 &$-0.22\pm 0.13$ 
&$26.21\pm0.15$& $>25.2$ & $>25.0$ & 2 \\
$  552/L21^e$       &13:38:33.56 & --19:43:36.00 &$-0.16\pm 0.18$
\ &$26.22\pm0.19$& $\dots$ & $\dots$ & 4 \\
$  507$       &13:38:34.26 & --19:43:12.20  &$0.10\pm 0.11$
&$26.22\pm0.12$& $25.21\pm0.44$ & $23.96\pm0.23$ & 3 \\
$  3564$       &13:38:23.34 & --19:41:51.49 &$-0.02\pm 0.23$
 &$26.22\pm0.29$& $25.15\pm0.42$ & $25.28\pm0.82$ & 3 \\
$  540$       &13:38:33.26 & --19:43:49.44  &$-0.08\pm 0.24$
&$26.28\pm0.26$& $\dots$ & $\dots$ & 4 \\
$  2480$       &13:38:27.25 & --19:42:30.42 &$-0.04\pm 0.14$ 
&$26.30\pm0.21$& $\dots$ & $\dots$ & 4 \\
$  2712$       &13:38:26.54 & --19:42:12.01 &$-0.10\pm 0.13$
 &$26.34\pm0.18$& $24.27\pm0.30$ & $24.88\pm0.68$ & 1 \\
$  2494$       &13:38:25.39 & --19:43:34.79 &$ -0.14\pm 0.13 
$ &$26.35\pm0.19$& $\dots$ & $\dots$ & 4 \\
$  2708$       &13:38:24.13 & --19:43:50.55 &$-0.04\pm 0.23$ 
&$26.39\pm0.22$& $>25.2$ & $>25.0$ & 2 \\
$  538/L20^e$       &13:38:32.83 & --19:44:06.93 &$ 0.13\pm 0.14 $
 &$26.44\pm0.16$& $>25.2$ & $>25.0$ & 2 \\
$  1843$       &13:38:29.53 & --19:42:38.81  &$-0.21\pm 0.17$
&$26.47\pm0.21$& $25.18\pm0.63$ & $24.50\pm0.47$ & 3 \\
$  1876$       &13:38:30.04 & --19:42:27.79  &$0.03\pm 0.21$
&$26.48\pm0.25$& $\dots$ & $\dots$ & 4 \\
$  375$       &13:38:32.71 & --19:44:38.30  &$0.09\pm 0.23$
&$26.49\pm0.27$& $\dots$ & $\dots$ & 4 \\
$  1655$       &13:38:29.52 & --19:43:10.60 &$0.27\pm 0.14$ 
&$26.51\pm0.18$& $\dots$ & $\dots$ & 4 \\
$  1339/L14^e$       &13:38:28.72 & --19:44:36.98 &$0.24\pm 0.18$
 &$26.52\pm0.18$& $>25.2$ & $>25.0$ & 2 \\
$  286$       &13:38:34.08 & --19:43:58.08  &$-0.20\pm 0.14$
&$26.52\pm0.19$& $25.23\pm0.47$ & $>25.0$ & 1 \\
$  3133$       &13:38:23.75 & --19:42:56.64 &$0.04\pm 0.19 $ 
&$26.53\pm0.27$&  $>25.2$ & $>25.0$  & 2 \\
$  1800$       &13:38:29.65 & --19:42:39.82 &$-0.01\pm 0.19$ 
&$26.56\pm0.26$& $\dots$ & $\dots$ & 4 \\
$  3486$       &13:38:21.49 & --19:43:21.67 &$-0.06\pm 0.16$ 
&$26.61\pm0.21$&  $>25.2$ & $>25.0$  & 2 \\
$  2874/L4^e$       &13:38:22.46 & --19:43:33.68 &$-0.14\pm 0.16$
 &$26.68\pm0.23$&  $\dots$ & $\dots$ & 4 \\
$  1211$       &13:38:33.53 & --19:42:09.19  &$0.19\pm 0.24$
&$26.72\pm0.27$&  $>25.2$ & $>25.0$  & 2 \\
$  1203$       &13:38:31.76 & --19:42:53.82  &$0.04\pm 0.15$
&$26.73\pm0.26$&  $\dots$ & $\dots$ & 4 \\
$  2571$       &13:38:23.70 & --19:44:32.20  &$-0.00\pm 0.25$
&$26.76\pm0.23$& $25.42\pm0.49$ & $24.58\pm0.45$ & 3 \\
$  1265$       &13:38:28.64 & --19:44:52:15  &$0.36\pm 0.19$
&$26.82\pm0.18$& $25.28\pm0.59$ & $24.39\pm0.38$ & 3 \\
$  1712$       &13:38:27.93 & --19:44:05:61  &$-0.26\pm 0.14$
&$26.83\pm0.31$& $25.59\pm0.80$ & $>25.0$ & 3 \\
$  3013$       &13:38:24.18 & --19:43:03:36  &$0.05\pm 0.21$
&$26.85\pm0.30$& $24.83\pm0.32$ & $>25.0$ & 1 \\
$  1866$       &13:38:29.33 & --19:42:44:08  &$0.24\pm 0.27$
&$26.86\pm0.30$&  $>25.2$ & $>25.0$  & 2 \\
$  1290$       &13:38:32.15 & --19:42:25:56  &$-0.05\pm 0.24$
&$26.88\pm0.29$& $24.72\pm0.37$ & $24.83\pm0.56$ & 1 \\
\enddata
\tablenotetext{a}{IDs refer to Overzier et al. (2008).}
\tablenotetext{b}{Kron magnitude.}  \tablenotetext{c}{$Spitzer$ IRAC
  3.6 $\mu$m and 4.5 $\mu$m magnitude from GALFIT.}
\tablenotetext{d}{The $Spitzer$ confusion classes: (1) isolated and
  detected; (2) isolated but undetected; (3) confused, but GALFIT may
  help; (4) heavily blended.}  \tablenotetext{e}{$\rm Ly\alpha$
  Emitters.}  \tablenotetext{f}{The 3.6 $\mu$m measurement most probably includes
  a large contribution from the \ha\ emission line halo of the radio
  galaxy. Applying a correction for \ha\ yields a 3.6 $\mu$m continuum
  magnitude of $21.2\pm0.1$ (see \S3.1 for details). This is the value plotted in Figs. \ref{fig:z4} and \ref{fig:z5}.}
\end{deluxetable}

\begin{deluxetable}{lllrcrrr}
\tabletypesize\small
\tablecolumns{9}
\tablewidth{0pc}
\tablecaption{\label{tab:laesz4}Photometric Properties of 
Spectroscopically Confirmed Ly$\alpha$ Emitters in the Field of TN J1338--1942.}
\tablehead{\multicolumn{1}{c}{ID$^a$} & \multicolumn{1}{c}
{$\alpha_{J2000}$} & \multicolumn{1}{c}{$\delta_{J2000}$} &
\multicolumn{1}{c}{$i_{775}-z_{850}$} & \multicolumn{1}{c}{$z_{850}^b$} & 
\multicolumn{1}{c}{$3.6 \mu m^c$} & \multicolumn{1}{c}{$4.5 \mu m^c$} &
 \multicolumn{1}{c}{Class$^d$} }
\startdata
$\rm  L4$       &13:38:22.46 & --19:43:33.68  &$-0.14\pm 0.16$
&$26.68\pm0.23$&  $\dots$ & $\dots$ & 4 \\
$\rm L7$         & 13 38 24.78 & --19 41 33.66    &$0.32\pm 0.25$
  &  $27.20\pm0.49$& $>25.2$ & $>25.0$ & 2 \\
$\rm L8$         & 13 38 24.86 & --19 41 45.49     &$-0.26\pm 0.23$
 &  $26.51\pm0.30$& $>25.2$ & $>25.0$ & 2 \\
$ \rm L9$       &13:38:25.09 & --19:43:10.77  &$ -0.02\pm 0.07$
&$25.34\pm0.08$& $24.02\pm0.15$&$24.37\pm0.34$ & 1 \\
$ \rm L11$       &13:38:26.16 & --19:43:34.31 &$-0.08\pm 0.09$ 
&$25.94\pm0.10$& $>25.2$&$>25.0$ & 2 \\
$ \rm L14$       &13:38:28.72 & --19:44:36.98 &$0.24\pm 0.18 $
 &$26.52\pm0.18$& $>25.2$ & $>25.0$ & 2 \\
$ \rm L16$       &13:38:29.66 & --19:43:59.87 &$ -0.00\pm 0.11$ 
&$25.54\pm0.16$& $24.29\pm0.24$ & $25.35\pm0.97$ & 3 \\
$ \rm L17$       &13:38:29.86 & --19:43:25.84 &$0.44\pm 0.29 $ 
&$27.37\pm0.28$& $>25.2$ & $>25.0$ & 2 \\
$ \rm L20$       &13:38:32.83 & --19:44:06.93 &$0.13\pm 0.14$ 
&$26.44\pm0.16$& $>25.2$ & $>25.0$ & 2 \\
$\rm L21$       &13:38:33.56 & --19:43:36.00  &$-0.16\pm 0.18$
 &$26.22\pm0.19$& $\dots$ & $\dots$ & 4 \\
$ \rm L22$       &13:38:34.13 & --19:42:52.68 &$0.13\pm 0.13$ 
&$26.64\pm0.14$& $24.75\pm0.45$ & $24.99\pm0.77$ & 3 \\
$ \rm L25$       &13:38:34.96 & --19:42:24.98 &$ 0.00\pm 0.12$
 &$25.81\pm0.15$& $>25.2$ &$>25.0$ &2 \\
\enddata
\tablenotetext{a}{IDs refer to Overzier et al. (2008).}
\tablenotetext{b}{Kron magnitude.}
\tablenotetext{c}{$Spitzer$ IRAC 3.6 $\mu$m and 4.5 $\mu$m 
magnitude from GALFIT.}
\tablenotetext{d}{The $Spitzer$ confusion classes: (1) isolated 
and detected; (2) isolated but undetected; (3) confused, but 
GALFIT may help;(4) heavily blended.}
\end{deluxetable}

\begin{deluxetable}{llllrcr}
\tablecolumns{9}
\tablewidth{0pc}
\tablecaption{\label{tab:lbgz5}Photometric Properties of $V_{606}$-Dropouts 
in the Field of TN J0924--2201.}
\tablehead{\multicolumn{1}{c}{ID$^a$} & \multicolumn{1}{c}{$\alpha_{J2000}$}
 & \multicolumn{1}{c}{$\delta_{J2000}$} & \multicolumn{1}{c}{$z_{850}^b$} &
 \multicolumn{1}{c}{$3.6 \mu m^c$} & \multicolumn{1}{c}{Class$^d$} }
\startdata
  $  1873$       &09:24:15.15 &     --22:01:52.8   &$24.21\pm0.04$&$22.95\pm0.12$& 3 \\
  $   119$       &09:24:29.04 &     --22:02:40.7  &$24.82\pm0.09$&$23.35\pm0.14$& 3 \\
  $   303 $      &09:24:29.01 &     --22:01:52.8  &$24.82\pm0.07$&$23.87\pm0.14$& 3 \\
  $   444 $      &09:24:28.10 &     --22:01:46.9  &$25.20\pm0.11$&$23.30\pm0.07$& 1 \\
  $   1814 $     &09:24:19.76 &     --22:59:57.7  &$25.29\pm0.09$&$25.23\pm0.27$& 3 \\
  $   310 $      &09:24:28.09 &     --22:02:17.8  &$25.42\pm0.08$&$25.55\pm0.43$& 1 \\
$1396$/RG  &09:24:19.89 &     --22:01:41.3   &$25.45\pm0.12$&$22.16\pm0.05$& 3\\
$    1979$       &09:24:12.50 &     --22:02:45.5  &$25.46\pm0.12$&$24.70\pm0.30$& 3 \\
$    1802 $      &09:24:14.90 &     --22:02:15.3  &$25.46\pm0.10$&$\dots$& 4 \\
$     871 $      &09:24:25.73 &     --22:01:11.1  &$25.50\pm0.13$&$24.40\pm0.26$& 3 \\
$     595 $      &09:24:26.98 &     --22:01:37.1  &$25.53\pm0.14$&$\dots$& 4 \\
$     894 $      &09:24:23.13 &     --22:02:16.2  &$25.53\pm0.09$&$23.77\pm0.12$& 3 \\
$    1047  $     &09:24:22.17 &     --22:01:59.2  &$25.65\pm0.15$&$23.67\pm0.06$& 1 \\
$449$ (2881)$^e$ &09:24:23.89 &     --22:03:44.4 &$25.80\pm0.09$&$\dots$& 4  \\
$    1736  $     &09:24:18.92 &     --22:00:42.2  &$25.92\pm0.12$&$24.80\pm0.21$& 3 \\
$     670 $      &09:24:25.38 &     --22:02:06.9  &$25.93\pm0.16$&$24.84\pm0.28$& 3 \\
$     739 $      &09:24:25.74 &     --22:01:42.6  &$26.09\pm0.15$&$25.58\pm0.44$& 3 \\
$    1074 $      &09:24:21.22 &     --22:02:21.2  &$26.17\pm0.18$&$\dots$& 4 \\
$     510 $      &09:24:28.79 &     --22:01:11.8  &$26.27\pm0.12$&$>25.5$& 2 \\
$    1385 $      &09:24:19.29 &     --22:02:01.4  &$26.30\pm0.15$&$24.71\pm0.50$& 3 \\
$1844$ (1388)$^e$&09:24:16.66 &     --22:01:16.4  &$26.33\pm0.17$&$\dots$& 4 \\
$     505  $     &09:24:25.42 &     --22:02:48.0  &$26.36\pm0.22$&$\dots$& 4 \\
$    1898  $     &09:24:14.15 &     --22:02:15.5  &$26.49\pm0.19$&$\dots$& 4 \\
$(2849)^e $          & 09:24:24.29 &    --22:02:30.11 & $27.06\pm0.25$&$>25.5$ & 2 \\
$(2688)^e$          & 09:24:25.65 &    --22:03:00.27 & $>28.53$&$\dots$ & 4 \\

\enddata
\tablenotetext{a}{IDs refer to Overzier et al. (2006).}
\tablenotetext{b}{Kron magnitude.}
\tablenotetext{c}{$Spitzer$ IRAC 3.6 $\mu$m magnitude from GALFIT.}
\tablenotetext{d}{The $Spitzer$ confusion classes: (1) isolated and detected; 
(2) isolated but undetected; (3) confused, but GALFIT may help;
(4) heavily blended.}
\tablenotetext{e}{ $\rm Ly\alpha$ Emitters.}
\end{deluxetable}

\begin{deluxetable}{clllrcrr}
\tablecolumns{9}
\tablewidth{0pc}
\tablecaption{\label{tab:qsoz6}Photometric Properties of $i_{775}$-Dropouts 
in the Field of SDSS0836+0054 and SDSSJ1030+0524.}
\tablehead{\multicolumn{1}{c}{ID$^a$} & \multicolumn{1}{c}{$\alpha_{J2000}$} 
& \multicolumn{1}{c}{$\delta_{J2000}$} & \multicolumn{1}{c}{$z_{850}^b$} & 
\multicolumn{1}{c}{$3.6 \mu m^c$} & \multicolumn{1}{c}{$4.5 \mu m^c$}  &
 \multicolumn{1}{c}{Class$^d$}}
\startdata
SDSS0836       & 08:36:43.87 & 00:54:53.15 & $18.85\pm0.02$ & $17.89\pm0.01$ & $\dots$ & 1 \\
$\rm A $          & 08:36:45.25 &    00:54:10.99 & $25.54\pm0.10$&$23.78\pm0.09$ & $\dots$&3 \\
$\rm B $          & 08:36:47.05 &    00:53:55.90 & $26.00\pm0.17$&$\dots$ & $\dots$&4  \\
$\rm C $          & 08:36:50.10 &    00:55:31.16 & $26.24\pm0.15$&$\dots$ & $\dots$&4  \\
$\rm D $          & 08:36:48.21 &    00:54:41.19 & $26.42\pm0.14$&$23.94\pm0.24$ & $\dots$ &3 \\
$\rm E $          & 08:36:44.03 &    00:54:32.79 & $26.39\pm0.16$&$\dots$ & $\dots$ &4  \\
$\rm F $          & 08:36:42.67 &    00:54:44.00 & $26.03\pm0.17$&$\dots$ & $\dots$ &4  \\
$\rm G $          & 08:36:45.96 &    00:54:40.53 & $26.36\pm0.25$&$23.77\pm0.15$ & $\dots$ & 3  \\
SDSS1030          & 10:30:27.09 & 05:24:55.00   & $20.09\pm0.01$ & $19.30\pm0.01$ &$\dots$ & 1 \\
$\rm A6$          & 10 30 22.66 &    05 24 37.16 & $26.09\pm0.19$&$\dots$ & $\dots$ &4  \\
$\rm A11$          & 10 30 20.62 &    05 23 43.63 & $26.11\pm0.17$&$\dots$ & $\dots$ &4  \\
$\rm A12 $          & 10:30:28.23 &   05:22:35.64 & $26.32\pm0.15 $&$>25.2$ & $>24.6$ & 2  \\
$\rm A13 $          & 10:30:24.08 &   05:24:20.40 & $25.90\pm0.15$&$>25.2$ & $>24.6$ &2  \\
$\rm A14 $          & 10:30:21.74 &   05:25:10.80 & $25.59\pm0.13$&$24.06\pm0.34$ & $23.97\pm0.48$ & 3 \\
\enddata
\tablenotetext{a}{IDs refer to Zheng et al. (2006) and Kim et al. (2008).}
\tablenotetext{b}{Kron magnitude.}
\tablenotetext{c}{$Spitzer$ IRAC 3.6 $\mu$m from GALFIT.}
\tablenotetext{d}{The $Spitzer$ confusion classes: (1) isolated and detected; (2) 
isolated but undetected; (3) confused, but GALFIT may help;
(4) heavily blended.}
\end{deluxetable}
 
\end{document}